\documentclass{aa}
\usepackage{graphics}

\newcommand{\sub}[1]{\mbox{$_{\rm #1}$}}

\newcommand{\feh}{\mbox{[Fe/H]}}

\newcommand{\Teff}{\mbox{$T\sub{eff}$}}

\newcommand{\diff}{\mbox{d}}
\newcommand{\comment}[1]{}
      \def\msol{$M_\odot$}
       \def\enh{$\alpha$-enhanced~}

\begin{document}

\thesaurus{ 06(08.05.3,08.09.3,08.08.1) }

\title{Evolutionary tracks and isochrones for $\alpha$-enhanced stars}

\author{Bernardo Salasnich$^1$, L\'eo Girardi$^{1,2}$, Achim Weiss$^2$ and 
           Cesare Chiosi$^{1}$ }
\institute{
    $^1$ Department of Astronomy, University of Padova,
      Vicolo dell'Osservatorio 5, 35122 Padova, Italy \\         
    $^2$ Max-Planck-Institut f\"ur Astrophysik,
        Karl-Schwarzschild-Str.~1, D-85740 Garching bei M\"unchen,
        Germany\\ }

\offprints{L\'eo Girardi \\ 
e-mail: lgirardi@pd.astro.it }

\date{Received {22.05.2000}. Accepted {14.07.2000}.}

\authorrunning{B. Salasnich, L. Girardi, A. Weiss \& C. Chiosi}
\titlerunning{$\alpha$-enhanced evolutionary tracks}
\maketitle

\begin{abstract} 
We present four large sets of evolutionary tracks for stars with 
initial chemical compositions 
$[Y=0.250, Z=0.008]$, $[Y=0.273, Z=0.019]$, $[Y=0.320, Z=0.040]$
and $[Y=0.390, Z=0.070]$ and enhancement of $\alpha$ elements with
respect to the solar pattern.
The major improvement with respect to previous similar calculations
is that we use consistent opacities -- i.e.
computed with the same chemical composition as adopted in the stellar 
models -- over the whole relevant range of temperatures.
For the same initial chemical compositions $[Y,Z]$ and otherwise identical
input physics  we  present also
new evolutionary sequences with 
solar-scaled  mixtures of abundances. {Based on} these stellar
models we calculate the corresponding sets of isochrones 
both in the Johnson-Cousins
$UBVRIJHK$ and HST/WFPC2 photometric systems. Furthermore, we 
derive integrated magnitudes,
colours and mass-to-light ratios for ideal single stellar populations 
with total mass equal to $1\, M_{\odot}$. Finally, the major 
changes in the
tracks, isochrones, and integrated magnitudes and colours
passing from solar-scaled to \enh mixtures are briefly outlined. 
Retrieval of the complete data set is possible via the www page
 \verb$http://pleiadi.pd.astro.it$.

\keywords{Stars: evolution - Stars: interiors - Stars: Hertzsprung-Russell 
diagram} 
\end{abstract}

\section{Introduction}

There are indications both from observations and  theoretical 
studies that  stars in many astrophysical environments may have 
 the relative abundances of elements synthesized by 
nuclear $\alpha$-capture reactions (e.g.\ C, O, Ne, Mg, Si, Ti, etc)
that are
enhanced with respect to the solar ones. In brief, 
King (1994) found the mean [O/Fe] ratio for a handful of halo stars 
to be $+0.52$~dex. Similar results are found for 
globular cluster stars (e.g.\ [$\alpha/{\rm Fe}]=0.18$~dex for 47 Tuc, 
Carney 1996) and galactic bulge K giants (McWilliam \& Rich 1994, 
Rich \& McWilliam 2000).
According to models of chemical evolution of the Galaxy, bulge stars 
should be $\alpha$-enhanced to reproduce the metallicity distribution 
(Matteucci \& Brocato 1990, Matteucci et al. 1999). 
Indications that the abundance 
ratios of elements in elliptical galaxies are not solar (e.g. 
$[{\rm Mg/Fe}] = 0.3 - 0.7$~dex) come from Worthey et al.\ (1992). 
Also synthetic metal line strengths 
(Weiss, Peletier \& Matteucci 1995) from high-metallicity models show 
that solar-scaled models do not fit the observed Mg2 (5180~\AA) and Fe 
(5270 and 5335~\AA) line indices.

Given these premises, it is of paramount importance to include
the enhancement of $\alpha$-elements 
in the population synthesis models designed  to
describe those galaxies. In the past this has been done  but using
models that suffered from many limitations. Firstly, till recently 
there were
no extended sets of opacities for $\alpha$-enhanced mixtures 
covering the complete relevant range of stellar temperatures.
Such opacity calculations are nowadays available.
Secondly, it has often been assumed that the behaviour of 
$\alpha$-enhanced isochrones can be easily mimicked by suitably 
scaling the metallicity \feh\ (Salaris et al.\ 1993). 
However, the validity of such a 
re-scaling has  only been {confirmed} for low metallicities, 
and it holds only for some particular cases like  all 
$\alpha$-elements being enhanced by the same factor.
Weiss et al.\ (1995) already pointed out the differences in 
the evolution due to $\alpha$-enhancement 
at solar and super-solar metallicities.
Recent work by Salaris \& Weiss (1998) {and VandenBerg et al.\ 
(2000)} also indicates that the 
simple re-scaling of the \feh\ content becomes risky at high
metallicities. Moreover, we should keep in mind that there is 
no {\em a priori} reason, from the point of view of both 
nucleosynthetic and chemical evolution theories, to expect
that all $\alpha$-elements are enhanced by the same factor.
From observations of halo and bulge stars, the degree of
enhancement of different $\alpha$-elements is, in general, not
constant with  differences amounting to $\pm0.2$~dex.

Therefore in this paper we intend to  re-analyze the 
impact of the enhancement of  $\alpha$-elements  on the properties of
metal-rich stellar populations (whose metallicity ranges from
half-solar to three times-solar). The analysis is made  
for a particular choice
of \enh abundance ratios, for which self-consistent opacities
are available. We start calculating extended sets
of stellar evolutionary tracks (Sect.~\ref{sec_tracks}), with
total metal contents $Z=0.008$, 0.019, 0.040 and 0.070  both for
solar-scaled (same abundance ratios as in the solar mix) 
and $\alpha$-enhanced patterns of abundances. 
Particular {attention is paid to the description of}
the adopted initial element
abundances and sources of opacity. {Based on} those stellar models
we derive large grids of isochrones and single stellar populations
for which we present magnitudes, colours and mass-to-light ratios in
many pass-bands of the Johnson and WFPC2 photometric systems
(Sect.~\ref{sec_isochrones}). Extensive
tabulations of stellar models, isochrones and photometric properties 
 are provided in electronic format for the purpose of general use.
Finally, a short summary is given in Sect.~\ref{sec_conclusion}. 

{
VandenBerg et al.\ (2000) recently presented extensive sets of 
evolutionary tracks for \enh mixtures, limited
to the range of low-mass stars ($0.5\le M/M_\odot\le1.0$). A constant 
enhancement of all $\alpha$-elements has been assumed (also in their
opacity tables). In the present work, instead, the choice of initial 
abundances has been guided by observational results. It is 
worth remarking that, within the present errors, none of both
choices seems to be strongly preferred.
}

\section {Stellar tracks}
\label{sec_tracks}

We calculate four grids of stellar models with initial masses from
0.15 to 20 $M_{\odot}$ and initial chemical compositions
$[Y=0.250, Z=0.008]$, $[Y=0.273, Z=0.019]$, $[Y=0.320, Z=0.040]$,
$[Y=0.390, Z=0.070]$.
The metallicities and the helium-to-metal enrichment ratio 
$\Delta Y/\Delta Z$ are chosen
in such a way to secure consistency with the Girardi et al.\ (2000) 
models. The  helium-to-metal enrichment law is $Y=0.23+2.25\,Z$.

For the above values of $[Y, Z]$, we compute both 
solar-scaled  and $\alpha$-enhanced sets of stellar tracks.
The evolutionary phases covered by the grids extend from 
the zero age main 
sequence (ZAMS) up to either the start of the 
{thermally pulsing asymptotic giant branch (TP-AGB) phase} or 
carbon ignition. 

\subsection {Input physics}

The input physics of the present stellar models is the same as in 
Girardi et al.\ (2000), apart from 
differences in (i) the  opacities, and (ii) 
the rates of energy loss by plasma neutrinos. 
A short description of the updated ingredients and the way
we deal with the enhancement of $\alpha$-elements is presented 
in the sections below.

\subsubsection{Initial chemical composition}

For a fixed total metallicity $Z$, the solar-scaled  abundance ratios of 
metals are taken from Grevesse \& Noels (1993). In the 
$\alpha$-enhanced case, we adopt a mixture in which only the 
elements resulting from nuclear $\alpha$-capture reactions (O,
Ne, Mg, Si, S, Ca, Ti, Ni) are enhanced with respect to the solar 
abundance ratios. This is the same mixture as 
used by Salaris et al.\ (1997) 
and Salaris \& Weiss (1998) to  study the age of the 
oldest metal-poor and disk metal-rich globular clusters. 
For the sake of clarity, Table \ref{elements} lists the 
abundances for   the solar-scaled and  $\alpha$-enhanced mixtures. 
Columns (2) and (4) show the abundance $A_{\rm el}$ of the elements in 
logarithmic scale, $A_{\rm el} = \log N_{\rm el}/N_{\rm H} + 12$, 
where $N_{\rm el}$ is the abundance by number. 
Columns (3) and (5) display the
abundance by mass fraction $X_{\rm el}$, relative to the metallicity $Z$.
Columns (6) and (7) list the enhancement ratio (relative  either to Fe or  
the total metal fraction) in spectroscopic notation. 
The enhancement factors come 
from the determinations of chemical abundances  in metal-poor field stars
by Ryan et al.\ (1991, cf.\ also Salaris \& Weiss 1998).
To derive the initial abundances of different isotopes we use the 
isotopic ratios (by mass) of Anders \& Grevesse (1989).

\begin{table*}
\caption{Abundance ratios in the adopted solar-scaled and $\alpha$-enhanced 
mixtures. See the text for details. }
\label{elements}
\begin{small}
\begin{center}
\begin{tabular}{c|cc|cccc}
\hline\noalign{\smallskip}
element  & \multicolumn{2}{c}{solar-scaled } & 
\multicolumn{4}{c}{$\alpha$-enhanced}  \\
      &   $A_{\rm el}$ & $X_{\rm el}/Z$ &  $A_{\rm el}$   &  
$X_{\rm el}/Z$ &   $[A_{\rm el}/{\rm Fe}]$ &  $[A_{\rm el}/{\rm M}]$   \\
 (1) & (2) & (3) & (4) & (5) & (6) & (7)  \\
\noalign{\smallskip}\hline\noalign{\smallskip}
   O  &   8.870  &  0.482273	&  9.370    &  0.672836  &   0.50  & $+$0.1442   \\
   Ne &   8.080  &  0.098668	&  8.370    &  0.084869  &   0.29  & $-$0.0658   \\
   Mg &   7.580  &  0.037573	&  7.980    &  0.041639  &   0.40  & $+$0.0441   \\
   Si &   7.550  &  0.040520	&  7.850    &  0.035669  &   0.30  & $-$0.0558   \\
   S  &   7.210  &  0.021142	&  7.540    &  0.019942  &   0.33  & $-$0.0258   \\
   Ca &   6.360  &  0.003734	&  6.860    &  0.005209  &   0.50  & $+$0.1441   \\
   Ti &   5.020  &  0.000211	&  5.650    &  0.000387  &   0.63  & $-$0.1875   \\
   Ni &   6.250  &  0.004459	&  6.270    &  0.002056  &   0.02  & $-$0.3371   \\
   C  &   8.550  &  0.173285    &  8.550    &  0.076451  &   0.00  & $-$0.3553   \\
   N  &   7.970  &  0.053152    &  7.970    &  0.023450  &   0.00  & $-$0.3553   \\
   Na &   6.330  &  0.001999    &  6.330    &  0.000882  &   0.00  & $-$0.3553   \\
   Cr &   5.670  &  0.001005    &  5.670    &  0.000443  &   0.00  & $-$0.3557   \\
   Fe &   7.500  &  0.071794    &  7.500    &  0.031675  &   0.00  & $-$0.3557   \\
\noalign{\smallskip}\hline
\end{tabular}
\end{center}
\end{small}
\end{table*}

\subsubsection{Opacities}

The  opacities we have adopted are the same as in Salaris, 
Degl'Innocenti \& Weiss (1997). They  are available at:
\verb$http://www-phys.llnl.gov/V_Div/OPAL/existing.html$.  
Details are given below. 

\begin{figure}
\resizebox{\hsize}{!}{\includegraphics{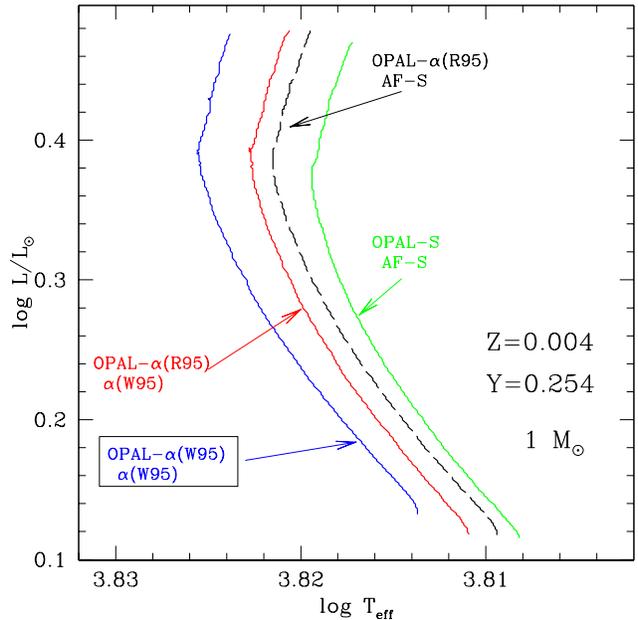}}
\caption{Evolution of a 1 $M_{\odot}$ star with the indicated
chemical composition  up to the  age of the Sun 
according to different 
opacity sources and  enhancements of $\alpha$-elements. 
See text for the meaning of the various labels and more details. }
\label{compare_opacity_hr}
\end{figure}

For high temperatures ($\log T \ge 4$) we use 
the radiative OPAL opacities by Rogers \& Iglesias (1995) 
in the case of solar-scaled  mixtures, and  by Iglesias \& Rogers (1996) 
for the \enh ones.
At low temperatures these are  combined with the 
molecular opacities provided by Alexander \& Ferguson 
(1994) for the solar mixtures, and by 
{Alexander (priv.\ communication)} for 
the \enh ones. In both cases, low- and high-temperature 
opacity grids  are merged by interpolation in the temperature
range from $\log T=3.8$ to $\log T= 4$.
For very high temperatures, $\log T \ge 8.7$, 
the Weiss et al.\ (1990) opacities are used. 
These rely on the Los Alamos opacity database. 
The radiative opacities  for C, O mixtures,
needed in the post-main sequence phases, are from 
Iglesias \& Rogers (1993).
Finally, conductive opacities of electron-degenerate 
matter are from Itoh et al.\ (1983).

It is very important to emphasize that the present 
models are based on {\em complete and consistent opacities}, 
i.e.\  the opacities (both for solar and \enh  mixtures) 
are generated for the same pattern of abundances 
 adopted in the stellar models, 
 both at low and high temperatures.

The importance of this issue is illustrated by 
Fig.~\ref{compare_opacity_hr}, which shows the effect in the
Hertzsprung-Russell (HR) diagram of adopting different (inconsistent) 
opacities and/or mixtures and/or enhancement factors as input to a 
1~$M_{\odot}$ main sequence model with $[Z=0.004, Y=0.254]$.
 Some of the illustrated cases are often found in  literature. 
The tracks are evolved up to the solar age of 4.6 Gyr. 
The meaning of the labels is the following: 
the first label refers to high temperature opacity ($\log T > 4$) 
and can vary from OPAL-S, i.e.\ solar-scaled  mixture, to OPAL-$\alpha$, 
i.e.\ enhanced according to Salaris \& Weiss (1998; W95) 
or {Rogers (priv. communication}; R95).
This latter  case  is a special $\alpha$-enhanced mixture characterized,
if compared with  Salaris \& Weiss (1998), by more C, less O, 
more Ne, more Na (which is not an alpha-element), less Al, more S, and 
more Fe.
The second label refers to the molecular opacity, with the same notation 
as for solar and $\alpha$-enhanced mixtures. 
AF stands for Alexander \& Ferguson (1994).  
The prescription enclosed in a box represents the one adopted by us.

Another point to emphasize is the difference between
the present opacities  and those used by 
Girardi et al.\ (2000). Two are the main 
sources of differences. First the adoption 
of the Itoh et al.\ (1983) conductive
opacities instead of the older ones by Hubbard \& Lampe (1969).
Interestingly, Catelan et al.\ (1996) show that 
Itoh et al.\ (1983) conductive opacities should not be 
used in RGB models, because they  are valid 
only for a  liquid phase, and not for the   physical 
conditions existing in the interiors of RGB stars. 
However, we verified that the effect of using the Itoh et al.
(1983) opacities 
instead of the Hubbard \& Lampe (1969) is to increase 
the core mass at the helium flash by 0.006~\msol\ (see also
Castellani et al.\ 2000 and Cassisi et al.\ 1998), which is
a negligible effect. 

The second difference lies in the numerical 
technique used to interpolate within the grids of the opacity tables. 
In this paper we use the two-dimensional bi-rational cubic 
damped-spline algorithm (see \mbox{Schlattl} \& Weiss 1998; and Sp\"ath 1973),
whereas  Girardi et al.\ (2000) adopted the smooth bi-parabolic spline 
interpolation (SSI) 
already introduced in our code  by Bressan et al.\ (1993). We 
verified that the differences amount to about a few 
percent.

\begin{figure}
\resizebox{\hsize}{!}{\includegraphics{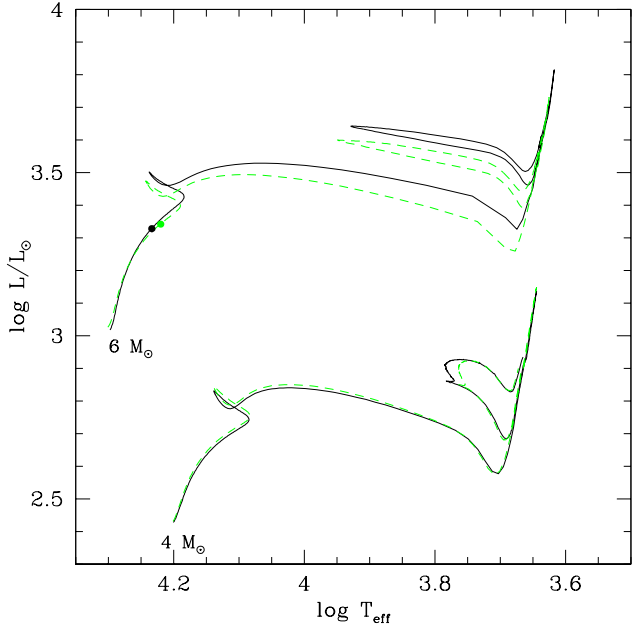}}
\caption{Evolution of the 4 and 6 \msol\ models according to 
different opacity interpolation schemes. The dashed lines
are  tracks calculated with the damped spline, whereas the 
continuous lines are  tracks calculated with 
SSI spline. The dots indicate the points when the central
hydrogen mass fraction is $X_{\rm c}=0.2$.  See also the companion 
Fig.~\protect\ref{compare_opacity_inner} for more details on the opacity.}
\label{compare_tracks_6msun}
\end{figure}

In general, evolutionary models of low mass differ very little by 
changing the opacity interpolation scheme. However,  massive stars 
(say $M \ge 5 M_{\odot}$) are much more sensitive
to the interpolation algorithm. This is shown 
in Fig.~\ref{compare_tracks_6msun},
where we plot two 4~\msol\ and 6~\msol\ models with $[Z=0.008, Y=0.250]$
calculated both with the SSI  
(continuous lines) and the damped-spline  
(dashed lines) scheme.

\begin{figure}
\resizebox{\hsize}{!}{\includegraphics{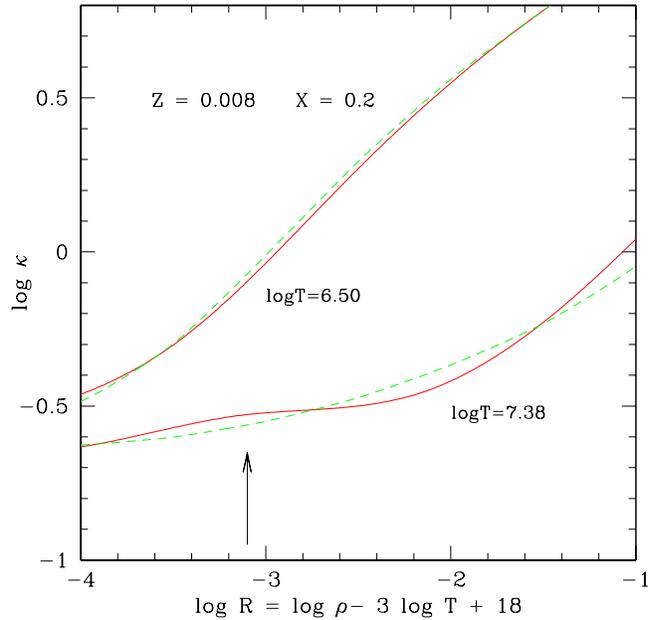}}
\caption{The opacity $\kappa$ as a function of the density and
temperature as indicated. 
The dashed lines are for damped-splined 
opacity interpolations, the continuous lines are for 
the bi-parabolic splined ones. The arrow shows the location 
of the Schwarzschild border  
of the convective core in a 6 \msol\ model with initial 
chemical composition $[Z=0.008, Y=0.250]$ during
the main sequence phase, when the central hydrogen mass 
fraction is $X_{\rm c}=0.2$. For this model the temperature 
at the core border is $\log T=7.38$.}
\label{compare_opacity_inner}
\end{figure}

Figure~\ref{compare_opacity_inner} compares  the effects of the different 
interpolations  on the opacity at the physical conditions holding 
in the inner regions of a 6~\msol\ star when the central hydrogen mass fraction
is $X_{\rm c}=0.2$, i.e.\ at the stage marked with a dot 
in Fig.~\ref{compare_tracks_6msun}.
At this stage the temperature at the Schwarzschild 
border of the convective core 
is $\log T =7.38$ and the density is $\log \rho = 1.05$
(corresponding to the arrow in the Fig.~\ref{compare_opacity_inner}).
For the  temperatures indicated along the curves, 
these represent the opacity 
as a function of the quantity $\log R=\log \rho - 3\log T + 18$ 
(see Rogers \& Iglesias 1992 for more details).
The continuous curve is based on the SSI interpolation, 
whereas the dashed curve refers to the method adopted in the present 
study. The small  difference in opacity at the core border  is typical of
the values found throughout the evolutionary sequences, 
and is responsible for the different HR patterns. 

We notice that both interpolations are acceptable, 
because the percentage error associated to both of them is
roughly 3 \%, which is lower than the opacity uncertainty 
in the original OPAL tables (the latter can be as high as 10 \% 
at individual $T/\rho$ points).
This shows that even the fine OPAL grid for some applications
is not sufficiently dense and more grid points might be needed.
This is also illustrated by the Solar Model by Schlattl \& Weiss (1998), 
where the inverse situation, i.e.\ spline producing a nicer
opacity run, is found.

The reason for the  different luminosities in the high-mass
models, is that in these stars the radiative gradient across the border of the
convective core is  shallow. Therefore, even slightly larger opacities 
at the Schwarzschild border may cause an increase of the
convective core and lead to higher luminosities.
This  trend already begins during the core H-burning phase and 
remains during the subsequent evolution.
The  effects are detectable 
starting from stars with mass larger than $6-10$~\msol\ 
(see Fig. \ref{compare_tracks_6msun}) 
and becomes dramatic over 60~\msol. 
It is worth recalling that the maximum mass
of our stellar models is 20~\msol.

\subsubsection{Neutrino losses}

Energy losses by neutrinos are from  Haft et al.\ 
(1994). Compared with the previous ones of Munakata et al.\ (1985)
 used by Girardi et al.\ (2000), neutrino cooling during the RGB is more
efficient. This causes an increase of 0.005~\msol\ in the core mass
at helium ignition. We have checked this on a 1~\msol\ model
with $Z= 0.019$. Differences of the same order have been found
by other authors (e.g.\ Haft et al.\ 1994; 
Cassisi et al.\ 1998).

\subsubsection{Remaining physical input}

The remaining physical input is the same as in Girardi et al.\ (2000), 
to whom we refer for further details. In summary, 
we follow the  evolution of H, $^3$He, $^4$He, $^{12}$C, 
$^{13}$C, $^{14}$N, $^{15}$N, $^{16}$O, $^{17}$O, $^{18}$O, $^{20}$Ne, 
$^{22}$Ne, Mg, according to the nuclear reaction rates
of Caughlan \& Fowler (1988) and  Landr\'e et al.\ (1990).  
For Mg we refer to the total content of $^{24}$Mg, $^{25}$Mg and 
$^{26}$Mg.

\begin{figure}
\resizebox{\hsize}{!}{\includegraphics{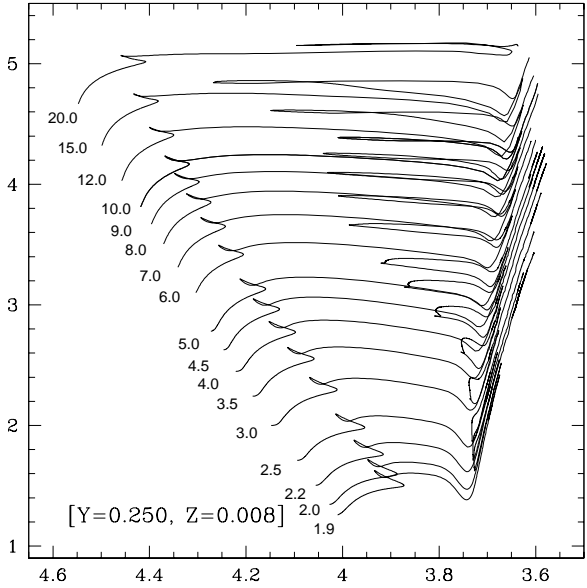}}
\caption{Evolutionary tracks for the composition $[Y=0.250, Z=0.008]$ 
and the $\alpha$-enhanced mixture of abundances.
The initial mass in solar units is indicated along each curve.}
\label{hrtracksz008}
\end{figure}

\begin{figure}
\resizebox{\hsize}{!}{\includegraphics{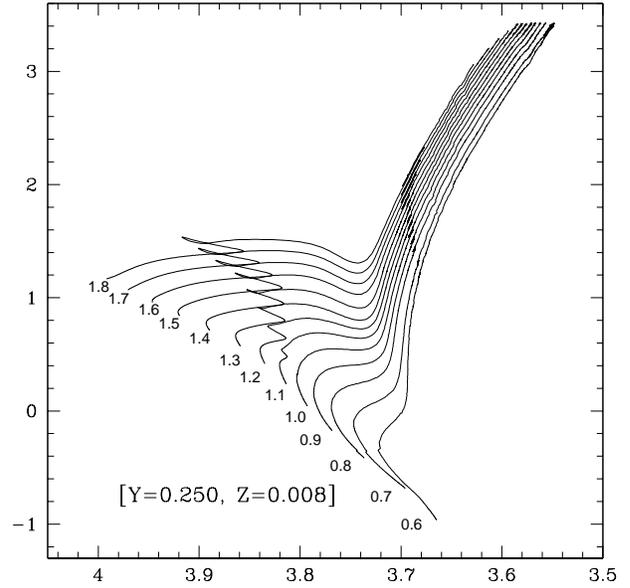}}
\caption{Evolutionary tracks (same composition as 
Fig.~\protect\ref{hrtracksz008}) for 
low-mass models up to the RGB-tip.}
\label{hrtracksz008_low}
\end{figure}

\begin{figure}
\resizebox{\hsize}{!}{\includegraphics{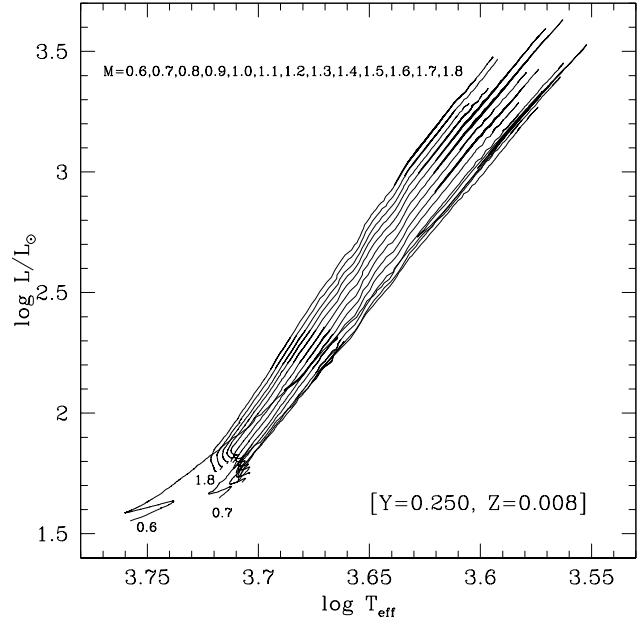}}
\caption{Evolutionary tracks (same composition as 
Fig.~\protect\ref{hrtracksz008}) for 
low-mass models from the ZAHB to the start of the TP-AGB phase.}
\label{hrtracksz008_low_hb}
\end{figure}

The equation of state (EOS) is that of 
a fully-ionized gas for temperatures higher that $10^7$~K. 
For lower temperatures we adopt the Mihalas et al.\ 
(1990 and references therein) EOS. 

Calibration of the solar model 
fixes the value of the  mixing length parameter $\alpha$  
of the B\"ohm-Vitense (1958) theory at 1.68.
Overshooting both from the convective core and  the
convective envelope are taken into account according to the 
 ballistic algorithm by Bressan et al.\ (1981).
Table \ref{overshooting} summarizes the  values of 
the overshooting parameters for the core, $\Lambda_{\rm c}$, 
and envelope, $\Lambda_{\rm e}$ we have adopted 
 as a function of the initial mass.

 As far as the mass-loss is concerned, this is taken into account  
and its effect on the stellar structure is calculated 
for masses higher than 6~\msol. We used the same prescription 
adopted by Fagotto et al.\ (1994), where more details 
can be found.

\begin{table}
\caption{Adopted parameters in the overshooting model by Bressan et al.\ 
(1981) and Alongi et al.\ (1991), as a function of the 
initial mass $M$ (in \msol).}
\label{overshooting}
\begin{center}
\begin{tabular}{c|c|c}
\hline\noalign{\smallskip}
     $M/M_\odot$  & $\Lambda_{\rm c}$  &  $\Lambda_{\rm e}$    \\
\noalign{\smallskip}\hline\noalign{\smallskip}
   0.5 - 1.0 & 0              & 0.25 \\
   1.1 - 1.4 & $M - 1$        & 0.25 \\
   1.5 - 2.0 & 0.5            & 0.25 \\
   2.0 - 2.5 & 0.5            & $0.5(2.5-M)+1.4(M-2)$  \\
   2.5 - 20  & 0.5            & 0.7 \\
\noalign{\smallskip}\hline
\end{tabular}
\end{center}
\end{table}

\subsection{HR diagram}

For the sake of illustration, in Figs.~\ref{hrtracksz008}, 
\ref{hrtracksz008_low} and \ref{hrtracksz008_low_hb} we present 
the complete set of evolutionary tracks for $[Y=0.250, Z=0.008]$ 
and $\alpha$-enhanced  mixtures. 

Fig.~\ref{hrtracksz008} shows the models of intermediate-mass 
stars from the zero age main sequence (ZAMS) up 
to the TP-AGB phase 
and those of massive stars from the ZAMS up to carbon ignition.
Fig.~\ref {hrtracksz008_low} presents the low-mass tracks 
from the ZAMS up to the RGB-tip. 
Finally, Fig.~\ref{hrtracksz008_low_hb} shows the corresponding 
He-burning phase from the zero age horizontal branch (ZAHB)
up to the TP-AGB phase.

{Some relevant} physical quantities are 
given in Table~\ref{timez008}, which lists, as a function of the 
initial mass, the lifetimes in years of the central H-burning 
and He-burning phases, $t_{\rm H}$ and $t_{\rm He}$ respectively, the core 
mass at the helium {ignition}, $M_{\rm c}({\rm HeF})$, and the core mass 
at the first thermal pulse on the AGB phase, $M_{\rm c}({\rm 1^{st}-TP})$.
For intermediate-mass and massive stars,   $M_{\rm c}({\rm HeF})$ 
 refers to  the core mass at the He-ignition.
For the same chemical composition Table~\ref{dredgeup} 
reports the changes in the surface chemical abundances 
of  those elements that suffered from nuclear processing
via the   pp-chain and CNO-cycle in the deep
interiors,  
after the completion of the first and the second dredge-up by envelope mixing.
The surface abundances of heavier elements are not altered 
by nuclear processing.
The last two columns give the ratios of 
$^{14}$N and $^{16}$O with respect to their initial 
values. These ratios are indicative of the efficiency of the dredge-up 
episodes. 

\begin{figure}
\resizebox{\hsize}{!}{\includegraphics{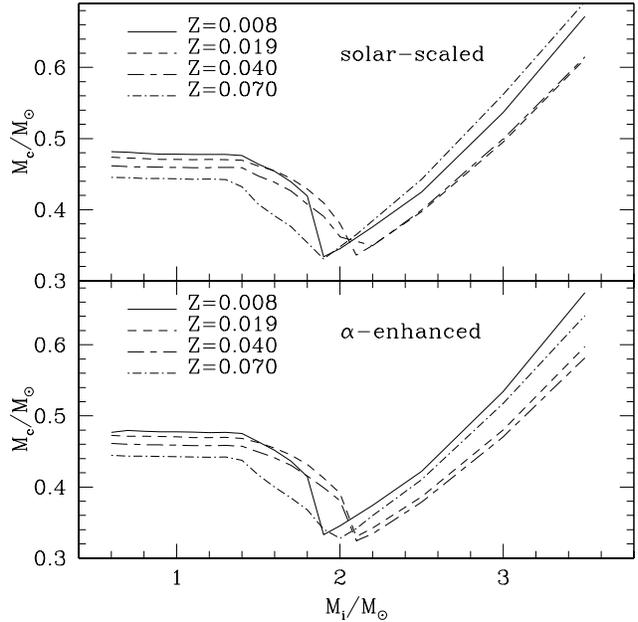}}
\caption{Core mass at He ignition as a function of the initial mass,
{for both solar-scaled (upper panel) and \enh\ (bottom panel) models.}}
\label{mflash}
\end{figure}

Table~\ref{critical_masses} displays the transition 
masses $M_{\rm conv}$, $M_{\rm HeF}$ and $M_{\rm up}$
for  models calculated with
solar and \enh initial composition. 
In brief, $M_{\rm conv}$ is the maximum mass below which stars become fully
convective during the central H-burning phase, $M_{\rm HeF}$ is the
maximum mass for the core He-flash to occur, and finally $M_{\rm up}$
is the maximum mass for the central C-burning to occur in highly
electron degenerate gas (deflagration or detonation followed by 
disruption of the star).

\begin{table*}
\caption{Lifetimes of the central H-burning $t_{\rm H}$ 
and He-burning $t_{\rm He}$ phases, the core mass at
the helium {ignition} $M_{\rm c}({\rm HeF})$ and the core mass 
$M_{\rm c}(\mbox{1-TP})$ at first thermal pulse on AGB phase, 
for the models with $Z=0.008$ and solar-scaled  mixture (columns 2 to 5)
and for the models with $Z=0.07$ and \enh  mixture (columns 6 to 9).}
\label{timez008}
\begin{center}
\begin{tabular}{c|cccc|cccc}
\hline\noalign{\smallskip}
& \multicolumn{4}{c}{$Z=0.008$ solar-scaled} & \multicolumn{4}{c}{$Z=0.07$ \enh } \\
\noalign{\smallskip} \cline{2-5} \cline{6-9} 
\noalign{\smallskip}
     $M/M_\odot$ & 
	 $t_{\rm H}/$yr  &  $t_{\rm He}/$yr   & $M_{\rm c}({\rm HeF})$ & $M_{\rm c}(\rm {1^{st}-TP})$ &
	 $t_{\rm H}/$yr  &  $t_{\rm He}/$yr   & $M_{\rm c}({\rm HeF})$ & $M_{\rm c}(\rm {1^{st}-TP})$ \\
\noalign{\smallskip}\hline\noalign{\smallskip}
  0.6 & 5.45~$10^{10}$ &   1.05~$10^{8}$ &  0.4814 &  0.5108 & 4.48~$10^{10}$ & 1.08~$10^8$ & 0.4445 &  0.5398 \\ 
  0.7 & 3.14~$10^{10}$ &   1.04~$10^{8}$ &  0.4806 &  0.5159 & 2.45~$10^{10}$ & 1.05~$10^8$ & 0.4434 &  0.5614 \\ 
  0.8 & 1.87~$10^{10}$ &   1.03~$10^{8}$ &  0.4793 &  0.5192 & 1.42~$10^{10}$ & 1.02~$10^8$ & 0.4434 &  0.5756 \\ 
  0.9 & 1.18~$10^{10}$ &   1.02~$10^{8}$ &  0.4781 &  0.5258 & 8.28~$10^9$    & 1.01~$10^8$ & 0.4430 &  0.5926 \\ 
   1.0 & 7.64~$10^{9}$ &   1.02~$10^{8}$ &  0.4782 &  0.5272 & 5.11~$10^9$    & 1.01~$10^8$ & 0.4426 &  0.5769 \\ 
   1.1 & 5.26~$10^{9}$ &   1.01~$10^{8}$ &  0.4776 &  0.5317 & 3.64~$10^9$    & 9.92~$10^7$ & 0.4424 &  0.6049 \\ 
   1.2 & 3.90~$10^{9}$ &   1.00~$10^{8}$ &  0.4778 &  0.5312 & 2.89~$10^9$    & 9.87~$10^7$ & 0.4418 &  0.6079 \\ 
   1.3 & 3.19~$10^{9}$ &   9.83~$10^{7}$ &  0.4778 &  0.5291 & 2.39~$10^9$    & 9.71~$10^7$ & 0.4421 &  0.6049 \\ 
   1.4 & 2.68~$10^{9}$ &   9.79~$10^{7}$ &  0.4763 &  0.5314 & 2.01~$10^9$    & 9.89~$10^7$ & 0.4375 &  0.6064 \\ 
   1.5 & 2.28~$10^{9}$ &   1.07~$10^{8}$ &  0.4642 &  0.5273 & 1.71~$10^9$    & 1.11~$10^8$ & 0.4169 &  0.6022 \\ 
   1.6 & 1.87~$10^{9}$ &   1.14~$10^{8}$ &  0.4539 &  0.5245 & 1.40~$10^9$    & 1.23~$10^8$ & 0.3999 &  0.5947 \\ 
   1.7 & 1.57~$10^{9}$ &   1.26~$10^{8}$ &  0.4391 &  0.5194 & 1.16~$10^9$    & 1.36~$10^8$ & 0.3845 &  0.5875 \\ 
   1.8 & 1.33~$10^{9}$ &   1.44~$10^{8}$ &  0.4192 &  0.5065 & 9.76~$10^8$    & 1.46~$10^8$ & 0.3680 &  0.5913 \\ 
   1.9 & 1.15~$10^{9}$ &   3.10~$10^{8}$ &  0.3337 &  0.4893 & 8.28~$10^8$    & 1.70~$10^8$ & 0.3406 &  0.5849 \\ 
   2.0 & 9.90~$10^{8}$ &   2.71~$10^{8}$ &  0.3451 &  0.4909 & 7.11~$10^8$    & 2.15~$10^8$ & 0.3281 &  0.5730 \\ 
   2.2 & 7.65~$10^{8}$ &   2.10~$10^{8}$ &  0.3757 &  0.5104 & 5.33~$10^8$    & 1.71~$10^8$ & 0.3598 &  0.6114 \\ 
   2.5 & 5.44~$10^{8}$ &   1.44~$10^{8}$ &  0.4246 &  0.5452 & 3.66~$10^8$    & 1.21~$10^8$ & 0.4104 &  0.6117 \\ 
   3.0 & 3.39~$10^{8}$ &   7.36~$10^{7}$ &  0.5367 &  0.6483 & 2.16~$10^8$    & 6.60~$10^7$ & 0.5178 &  0.6844 \\ 
   3.5 & 2.31~$10^{8}$ &   3.98~$10^{7}$ &  0.6718 &  0.7806 & 1.40~$10^8$    & 3.77~$10^7$ & 0.6408 &  0.6844 \\ 
   4.0 & 1.67~$10^{8}$ &   2.48~$10^{7}$ &  0.8213 &  0.9375 & 9.71~$10^7$    & 2.34~$10^7$ & 0.7713 &  0.6844 \\ 
   4.5 & 1.27~$10^{8}$ &   1.67~$10^{7}$ &  0.9744 &  1.0910 & 7.12~$10^7$    & 1.56~$10^7$ & 0.9092 &  0.6844 \\ 
   5.0 & 1.00~$10^{8}$ &   1.21~$10^{7}$ &  1.1333 &  1.2497 & 5.45~$10^7$    & 1.10~$10^7$ & 1.0593 &  1.2062 \\ 
   6.0 & 6.72~$10^{7}$ &   6.73~$10^{6}$ &  1.4456 &   -     & 3.53~$10^7$    & 6.06~$10^6$ & 1.3904 &  1.5452 \\ 
   7.0 & 4.87~$10^{7}$ &   4.55~$10^{6}$ &  1.7673 &   -     & 2.51~$10^7$    & 3.89~$10^6$ & 1.7609 &   -     \\
   8.0 & 3.76~$10^{7}$ &   3.30~$10^{6}$ &  2.1265 &   -     & 1.91~$10^7$    & 2.74~$10^6$ & 2.1599 &   -     \\
   9.0 & 3.03~$10^{7}$ &   2.52~$10^{6}$ &  2.4985 &   -     & 1.53~$10^7$    & 2.06~$10^6$ & 2.5988 &   -     \\
  10.0 & 2.51~$10^{7}$ &   2.02~$10^{6}$ &  2.9138 &   -     & 1.28~$10^7$    & 1.62~$10^6$ & 3.0935 &   -     \\
  12.0 & 1.86~$10^{7}$ &   1.43~$10^{6}$ &  3.7804 &   -     & 9.59~$10^6$    & 1.14~$10^6$ & 3.9767 &   -     \\
  15.0 & 1.34~$10^{7}$ &   1.03~$10^{6}$ &  4.6706 &   -     & 7.10~$10^6$    & 8.26~$10^5$ & 5.3587 &   -     \\
  20.0 & 9.39~$10^{6}$ &   7.12~$10^{5}$ &  6.7187 &   -     & 5.17~$10^6$    & 6.01~$10^5$ & 7.7582 &   -     \\
\noalign{\smallskip}\hline									 \end{tabular}
\end{center}
\end{table*}

$M_{\rm HeF}$ is {here defined}
as the initial mass for which the core mass at He ignition has its
minimum value (see Table~\ref{timez008} and Fig.~\ref{mflash}). 
{Fig.~\ref{mflash} shows that similar values for $M_{\rm HeF}$ are found 
both in the \enh and in the solar-scaled case.}
Compared  with Girardi et al.\ (2000), we find 
higher $M_{\rm up}$ values. This is once more caused by the 
different interpolation schemes of the opacity tables. 
The  one currently in use yields smaller convective cores for 
the same initial mass.
It follows that with the present overshooting prescription 
the maximum mass for a star to develop an electron-degenerate 
core after the core He-burning  phase is slightly higher than 
previously estimated.

\begin{table*}
\caption{Surface chemical composition 
(by mass fraction) of the $[Z=0.008, Y=0.250]$ solar-scaled  
models.}
\label{dredgeup}
\begin{scriptsize}
\begin{tabular}{ccccccccccccc}
\noalign{\smallskip}\hline\noalign{\smallskip}
$M/M_{\odot}$ &  H  & $^3$He & $^4$He  & $^{12}$C & $^{13}$C & $^{14}$N & $^{15}$N & $^{16}$O & $^{17}$O & $^{18}$O 
  & $\frac{ ^{12}{\rm C} }{ ^{12}{\rm C}_{\rm i}}$  &  $\frac{ ^{14}{\rm N} }{ ^{14}{\rm N}_{\rm i} }$ \\
\noalign{\smallskip}\hline\noalign{\smallskip}
\multicolumn{11}{l}{Initial:}   \\
  all   & 0.742 &  3.80$\,10^{-5}$ &   0.250 &  1.37$\,10^{-3}$ &  1.65$\,10^{-5}$ &  4.24$\,10^{-4}$ &  1.67$\,10^{-6}$ &  3.85$\,10^{-3}$ &  1.56$\,10^{-6}$ &  8.68$\,10^{-6}$ & 1.00 & 1.00 \\
\noalign{\smallskip}\hline\noalign{\smallskip}
\multicolumn{11}{l}{After the first dredge-up:} \\
  0.60 &  0.730  & 3.95$\,10^{-3}$  &  0.258  & 1.37$\,10^{-3}$  & 1.72$\,10^{-5}$  & 4.24$\,10^{-4}$ &  1.66$\,10^{-6}$  
& 3.85$\,10^{-3}$  & 1.56$\,10^{-6}$  & 8.68$\,10^{-6}$  & 1.00  & 1.00 \\
  0.70 &  0.727  & 2.73$\,10^{-3}$  &  0.263  & 1.36$\,10^{-3}$  & 2.50$\,10^{-5}$  & 4.26$\,10^{-4}$ &  1.55$\,10^{-6}$  
& 3.85$\,10^{-3}$  & 1.56$\,10^{-6}$  & 8.68$\,10^{-6}$  & 0.993  & 1.01 \\
  0.80 &  0.725  & 1.97$\,10^{-3}$  &  0.265  & 1.32$\,10^{-3}$  & 3.71$\,10^{-5}$  & 4.59$\,10^{-4}$ &  1.44$\,10^{-6}$  
& 3.85$\,10^{-3}$  & 1.56$\,10^{-6}$  & 8.64$\,10^{-6}$  & 0.964  & 1.08 \\
  0.90 &  0.724  & 1.55$\,10^{-3}$  &  0.267  & 1.27$\,10^{-3}$  & 4.02$\,10^{-5}$  & 5.18$\,10^{-4}$ &  1.36$\,10^{-6}$  
& 3.85$\,10^{-3}$  & 1.57$\,10^{-6}$  & 8.52$\,10^{-6}$  & 0.925  & 1.22 \\
  1.00 &  0.723  & 1.20$\,10^{-3}$  &  0.267  & 1.22$\,10^{-3}$  & 4.13$\,10^{-5}$  & 5.72$\,10^{-4}$ &  1.30$\,10^{-6}$  
& 3.85$\,10^{-3}$  & 1.60$\,10^{-6}$  & 8.30$\,10^{-6}$  & 0.891  & 1.35 \\
  1.10 &  0.723  & 1.01$\,10^{-3}$  &  0.268  & 1.17$\,10^{-3}$  & 4.24$\,10^{-5}$  & 6.35$\,10^{-4}$ &  1.22$\,10^{-6}$  
& 3.85$\,10^{-3}$  & 1.70$\,10^{-6}$  & 8.04$\,10^{-6}$  & 0.851  & 1.50 \\
  1.20 &  0.725  & 8.67$\,10^{-4}$  &  0.267  & 1.14$\,10^{-3}$  & 4.38$\,10^{-5}$  & 6.70$\,10^{-4}$ &  1.18$\,10^{-6}$  
& 3.85$\,10^{-3}$  & 1.83$\,10^{-6}$  & 7.89$\,10^{-6}$  & 0.829  & 1.58 \\
  1.30 &  0.726  & 8.00$\,10^{-4}$  &  0.265  & 1.11$\,10^{-3}$  & 4.18$\,10^{-5}$  & 7.00$\,10^{-4}$ &  1.15$\,10^{-6}$  
& 3.85$\,10^{-3}$  & 2.01$\,10^{-6}$  & 7.72$\,10^{-6}$  & 0.811  & 1.65 \\
  1.40 &  0.724  & 6.89$\,10^{-4}$  &  0.268  & 1.02$\,10^{-3}$  & 4.59$\,10^{-5}$  & 8.06$\,10^{-4}$ &  1.02$\,10^{-6}$  
& 3.85$\,10^{-3}$  & 2.56$\,10^{-6}$  & 7.28$\,10^{-6}$  & 0.742  & 1.90 \\
  1.50 &  0.723  & 6.11$\,10^{-4}$  &  0.268  & 9.81$\,10^{-4}$  & 4.49$\,10^{-5}$  & 8.53$\,10^{-4}$ &  9.72$\,10^{-7}$  
& 3.84$\,10^{-3}$  & 1.02$\,10^{-5}$  & 7.02$\,10^{-6}$  & 0.716  & 2.01 \\
  1.60 &  0.724  & 5.35$\,10^{-4}$  &  0.268  & 9.56$\,10^{-4}$  & 4.47$\,10^{-5}$  & 9.02$\,10^{-4}$ &  9.44$\,10^{-7}$  
& 3.82$\,10^{-3}$  & 1.02$\,10^{-5}$  & 6.86$\,10^{-6}$  & 0.697  & 2.13 \\
  1.70 &  0.722  & 4.76$\,10^{-4}$  &  0.269  & 9.34$\,10^{-4}$  & 4.50$\,10^{-5}$  & 9.66$\,10^{-4}$ &  9.15$\,10^{-7}$  
& 3.77$\,10^{-3}$  & 1.70$\,10^{-5}$  & 6.76$\,10^{-6}$  & 0.681  & 2.28 \\
  1.80 &  0.720  & 4.21$\,10^{-4}$  &  0.272  & 9.04$\,10^{-4}$  & 4.39$\,10^{-5}$  & 1.04$\,10^{-3}$ &  8.88$\,10^{-7}$  
& 3.72$\,10^{-3}$  & 1.93$\,10^{-5}$  & 6.60$\,10^{-6}$  & 0.660  & 2.46 \\
  1.90 &  0.718  & 3.78$\,10^{-4}$  &  0.274  & 9.01$\,10^{-4}$  & 4.45$\,10^{-5}$  & 1.09$\,10^{-3}$ &  8.76$\,10^{-7}$  
& 3.67$\,10^{-3}$  & 1.63$\,10^{-5}$  & 6.55$\,10^{-6}$  & 0.658  & 2.57 \\
  2.00 &  0.718  & 3.38$\,10^{-4}$  &  0.274  & 8.95$\,10^{-4}$  & 4.41$\,10^{-5}$  & 1.12$\,10^{-3}$ &  8.72$\,10^{-7}$  
& 3.64$\,10^{-3}$  & 1.63$\,10^{-5}$  & 6.52$\,10^{-6}$  & 0.653  & 2.65 \\
  2.20 &  0.714  & 2.76$\,10^{-4}$  &  0.278  & 8.88$\,10^{-4}$  & 4.51$\,10^{-5}$  & 1.20$\,10^{-3}$ &  8.57$\,10^{-7}$  
& 3.57$\,10^{-3}$  & 1.53$\,10^{-5}$  & 6.45$\,10^{-6}$  & 0.648  & 2.82 \\
  2.50 &  0.711  & 2.13$\,10^{-4}$  &  0.280  & 8.72$\,10^{-4}$  & 4.53$\,10^{-5}$  & 1.27$\,10^{-3}$ &  8.37$\,10^{-7}$  
& 3.51$\,10^{-3}$  & 1.16$\,10^{-5}$  & 6.36$\,10^{-6}$  & 0.636  & 3.00 \\
  3.00 &  0.713  & 1.50$\,10^{-4}$  &  0.279  & 8.65$\,10^{-4}$  & 4.47$\,10^{-5}$  & 1.30$\,10^{-3}$ &  8.27$\,10^{-7}$  
& 3.48$\,10^{-3}$  & 9.83$\,10^{-6}$  & 6.30$\,10^{-6}$  & 0.631  & 3.08 \\
  3.50 &  0.717  & 1.13$\,10^{-4}$  &  0.275  & 8.78$\,10^{-4}$  & 4.66$\,10^{-5}$  & 1.27$\,10^{-3}$ &  8.23$\,10^{-7}$  
& 3.50$\,10^{-3}$  & 6.74$\,10^{-6}$  & 6.42$\,10^{-6}$  & 0.641  & 2.99 \\
  4.00 &  0.718  & 9.02$\,10^{-5}$  &  0.274  & 8.80$\,10^{-4}$  & 4.83$\,10^{-5}$  & 1.27$\,10^{-3}$ &  8.10$\,10^{-7}$  
& 3.50$\,10^{-3}$  & 5.56$\,10^{-6}$  & 6.43$\,10^{-6}$  & 0.642  & 2.99 \\
  4.50 &  0.720  & 7.69$\,10^{-5}$  &  0.272  & 9.13$\,10^{-4}$  & 4.77$\,10^{-5}$  & 1.22$\,10^{-3}$ &  8.49$\,10^{-7}$  
& 3.51$\,10^{-3}$  & 3.89$\,10^{-6}$  & 6.64$\,10^{-6}$  & 0.666  & 2.89 \\
  5.00 &  0.721  & 6.49$\,10^{-5}$  &  0.270  & 9.49$\,10^{-4}$  & 4.64$\,10^{-5}$  & 1.18$\,10^{-3}$ &  8.90$\,10^{-7}$  
& 3.51$\,10^{-3}$  & 3.70$\,10^{-6}$  & 6.73$\,10^{-6}$  & 0.693  & 2.79 \\
  6.00 &  0.722  & 4.75$\,10^{-5}$  &  0.270  & 8.75$\,10^{-4}$  & 4.96$\,10^{-5}$  & 1.28$\,10^{-3}$ &  7.79$\,10^{-7}$  
& 3.50$\,10^{-3}$  & 4.04$\,10^{-6}$  & 6.40$\,10^{-6}$  & 0.638  & 3.01 \\
  7.00 &  0.723  & 3.85$\,10^{-5}$  &  0.269  & 8.55$\,10^{-4}$  & 5.00$\,10^{-5}$  & 1.32$\,10^{-3}$ &  7.51$\,10^{-7}$  
& 3.47$\,10^{-3}$  & 3.84$\,10^{-6}$  & 6.28$\,10^{-6}$  & 0.624  & 3.12 \\
  8.00 &  0.721  & 3.26$\,10^{-5}$  &  0.271  & 8.41$\,10^{-4}$  & 4.98$\,10^{-5}$  & 1.37$\,10^{-3}$ &  7.31$\,10^{-7}$  
& 3.43$\,10^{-3}$  & 3.61$\,10^{-6}$  & 6.15$\,10^{-6}$  & 0.614  & 3.24 \\
  9.00 &  0.722  & 3.06$\,10^{-5}$  &  0.270  & 8.83$\,10^{-4}$  & 4.99$\,10^{-5}$  & 1.31$\,10^{-3}$ &  7.83$\,10^{-7}$  
& 3.45$\,10^{-3}$  & 2.77$\,10^{-6}$  & 6.42$\,10^{-6}$  & 0.644  & 3.08 \\
 10.00 &  0.721  & 2.78$\,10^{-5}$  &  0.271  & 8.79$\,10^{-4}$  & 4.92$\,10^{-5}$  & 1.33$\,10^{-3}$ &  7.79$\,10^{-7}$  
& 3.43$\,10^{-3}$  & 2.83$\,10^{-6}$  & 6.37$\,10^{-6}$  & 0.641  & 3.13 \\
 12.00 &  0.713  & 2.36$\,10^{-5}$  &  0.279  & 8.62$\,10^{-4}$  & 5.03$\,10^{-5}$  & 1.44$\,10^{-3}$ &  7.52$\,10^{-7}$  
& 3.33$\,10^{-3}$  & 2.49$\,10^{-6}$  & 6.21$\,10^{-6}$  & 0.629  & 3.40 \\
 15.00 &  0.687  & 1.90$\,10^{-5}$  &  0.305  & 8.23$\,10^{-4}$  & 4.94$\,10^{-5}$  & 1.67$\,10^{-3}$ &  7.15$\,10^{-7}$  
& 3.12$\,10^{-3}$  & 2.24$\,10^{-6}$  & 5.81$\,10^{-6}$  & 0.601  & 3.94 \\
\noalign{\smallskip}\hline\noalign{\smallskip} 
\multicolumn{11}{l}{After the second dredge-up:} \\
  4.00 &  0.694  & 8.64$\,10^{-5}$  &  0.298  & 8.46$\,10^{-4}$  & 4.71$\,10^{-5}$  & 1.42$\,10^{-3}$ &  7.84$\,10^{-7}$  
& 3.37$\,10^{-3}$  & 5.91$\,10^{-6}$  & 6.17$\,10^{-6}$  & 0.618  & 3.35 \\
  4.50 &  0.676  & 6.95$\,10^{-5}$  &  0.316  & 8.35$\,10^{-4}$  & 4.65$\,10^{-5}$  & 1.51$\,10^{-3}$ &  7.72$\,10^{-7}$  
& 3.29$\,10^{-3}$  & 4.59$\,10^{-6}$  & 6.07$\,10^{-6}$  & 0.609  & 3.56 \\
  5.00 &  0.662  & 5.71$\,10^{-5}$  &  0.330  & 8.48$\,10^{-4}$  & 4.51$\,10^{-5}$  & 1.56$\,10^{-3}$ &  7.90$\,10^{-7}$  
& 3.22$\,10^{-3}$  & 4.24$\,10^{-6}$  & 6.01$\,10^{-6}$  & 0.619  & 3.68 \\
  6.00 &  0.667  & 4.35$\,10^{-5}$  &  0.325  & 8.08$\,10^{-4}$  & 4.76$\,10^{-5}$  & 1.60$\,10^{-3}$ &  7.29$\,10^{-7}$  
& 3.22$\,10^{-3}$  & 3.82$\,10^{-6}$  & 5.86$\,10^{-6}$  & 0.590  & 3.77 \\
  7.00 &  0.714  & 3.75$\,10^{-5}$  &  0.278  & 8.35$\,10^{-4}$  & 4.89$\,10^{-5}$  & 1.41$\,10^{-3}$ &  7.38$\,10^{-7}$  
& 3.40$\,10^{-3}$  & 3.89$\,10^{-6}$  & 6.11$\,10^{-6}$  & 0.609  & 3.32 \\ 
\noalign{\smallskip}\hline\noalign{\smallskip} 	
\end{tabular}							    
\end{scriptsize}
\end{table*}

\begin{table}
\caption{The transition masses $M_{\rm conv}$, $M_{\rm HeF}$ and 
$M_{\rm up}$.}
\label{critical_masses}
\begin{center}
\begin{tabular}{c|c|c|c|c|c}
\hline\noalign{\smallskip}
     $Z$  & $Y$ & mixture  & $M_{\rm conv}$  & $M_{\rm HeF}$  &  
$M_{\rm up}$ \\
\noalign{\smallskip}\hline\noalign{\smallskip}
  0.008 & 0.250  &  solar-scaled  & 0.36 & 1.90 & $5-6$ \\ 
  0.019 & 0.273  &  solar-scaled  & 0.37 & 2.10 & $6-7$ \\
  0.040 & 0.320  &  solar-scaled  & 0.36 & 2.20 & $7-8$ \\
  0.070 & 0.390  &  solar-scaled  & 0.33 & 1.90 & $7-8$  \\ \hline
  0.008 & 0.250  &  \enh          & 0.37 & 1.90 & $5-6$ \\
  0.019 & 0.273  &  \enh          & 0.37 & 2.10 & $5-6$ \\
  0.040 & 0.320  &  \enh          & 0.36 & 2.10 & $7-8$ \\
  0.070 & 0.390  &  \enh          & 0.33 & 2.00 & $6-7$  \\
\noalign{\smallskip}\hline
\end{tabular}
\end{center}
\end{table}

\section {Isochrones}
\label{sec_isochrones}

The four sets of  stellar models are used to calculate  isochrones,  
integrated magnitudes and colours 
of single stellar populations with ages from  
$10^{7}$ to $2\times10^{10}$~yr
both in the Johnson-Cousins $UBVRIJHK$ and HST-WFPC2 photometric 
systems. The age range is large enough to describe
young clusters and associations as well as old globular clusters.
The dense grid of stellar tracks allows us to construct detailed
isochrones at small age steps ($\Delta \log t=0.05$). 
This also makes the data-base {suitable for the simulation of} 
synthetic colour-magnitude diagrams (e.g.\ Girardi 1999).

{Before theoretical isochrones are constructed, the TP-AGB phase 
is included in all tracks of $M<M_{\rm up}$. Therefore, this phase
is present in the isochrones older than about $10^8$~yr. 
The reader is referred to Girardi \& Bertelli (1998) and 
Girardi et al.\ (2000) for all details about the synthetic 
algorithm used to follow the TP-AGB evolution. }

\subsection{Isochrones in the HR-diagram} 

\begin{figure}
\resizebox{\hsize}{!}{\includegraphics{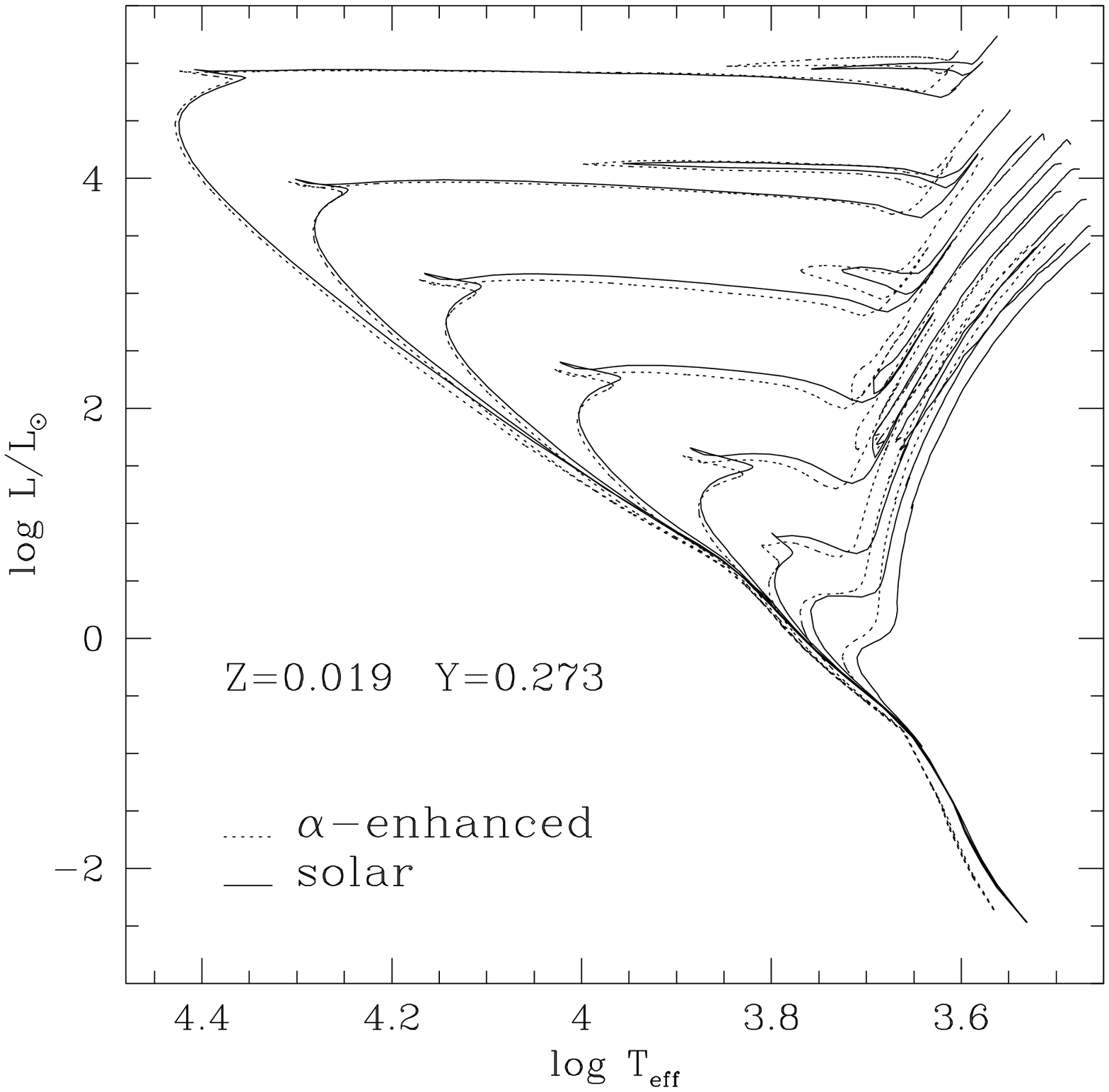}}
\caption{Comparison of isochrones with chemical composition 
$[Z=0.019, Y=0.273]$ and different enhancement of $\alpha$-elements.
Dotted lines refer to the \enh isochrones, continuous lines 
to the solar-scaled  ones. Only the isochrones with ages
between  $\log(t/{\rm yr}) = 7.0$ and $\log(t/{\rm yr}) =10.5$, 
in steps of  $\Delta \log t=0.5$ for the sake of clarity, are plotted.}
\label{compare_isoc}
\end{figure}

\begin{figure}
\resizebox{\hsize}{!}{\includegraphics{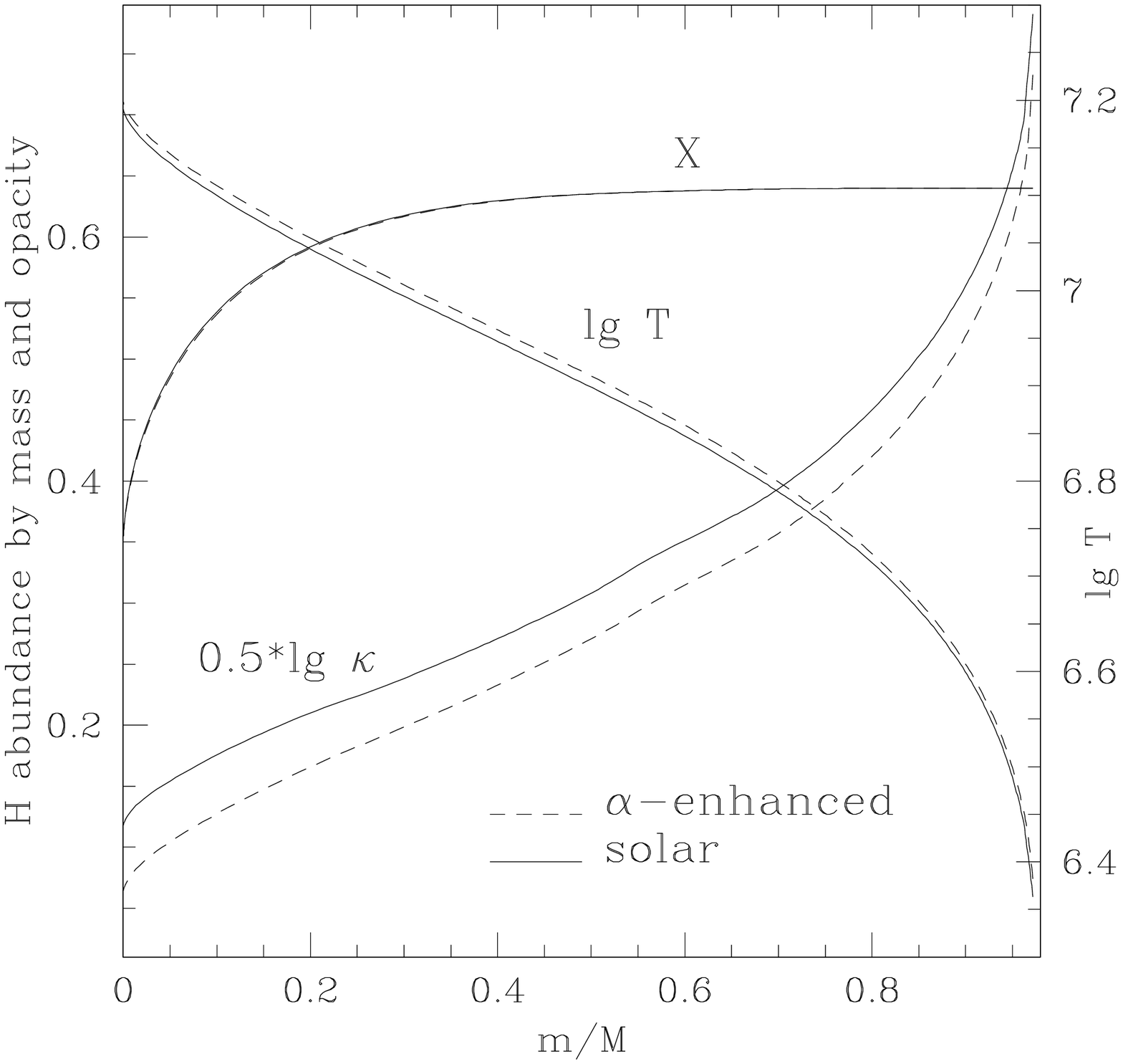}}
\caption{Comparison of the internal structure of two 1~\msol\ models with 
$[Z=0.040, Y=0.320]$ at the same evolutionary stage, i.e.\ when the central 
hydrogen mass fraction is 0.36. The left scale refers both to the hydrogen
abundance by mass and  the opacity. The right scale refers 
to the logarithm of the temperature. Continuous lines are 
for  the solar-scaled  model, whereas the dashed lines are for the \enh one.}
\label{compare_inner}
\end{figure}
The HR diagram of Fig.~\ref{compare_isoc} shows a comparison 
between isochrones with 
chemical composition $[Z=0.019, Y=0.273]$  both for solar-scaled  
(continuous lines) and \enh (dotted lines) stellar models. 
For the sake of clarity we plot only the isochrones with ages 
between  $\log(t/{\rm yr}) = 7.0$ and $\log (t/{\rm yr}) =10.5$, 
in steps of  $\Delta \log t=0.5$. 
As already noted by  Salaris \& Weiss (1998), for 
solar and super-solar total metallicity, the element 
abundance ratios among metals  affect the
evolution. From the HR diagram of Fig.~\ref{compare_isoc},
two main features can be singled out.

(i) $\alpha$-Enhanced  isochrones have  fainter and hotter turn-offs (TO).
This is shown by the entries of Table~\ref{turnoff}, which 
lists  the turn-off luminosity and effective temperature  
of isochrones with $Z=0.019$ but different mixtures of $\alpha$-elements.
The reason is that  at  given central 
hydrogen content the  stellar model with solar-scaled composition
has a mean opacity higher than  the  \enh one. The point is
illustrated in Fig.~\ref{compare_inner} where we compare the 
opacity profile across {the 1~\msol\ models} 
with $[Z=0.040, Y=0.320]$ and both \enh\  
(dashed line) and  solar-scaled   (solid line) mixtures. 
Higher opacities induce steeper radiative temperature gradients 
$\nabla_{r}$ and lower surface temperatures 
$T_{\rm eff}$ for the same central conditions. 
Furthermore, steeper $\nabla_{r}$s  imply smaller
burning regions and lower luminosities in turn. 
This effect overwhelms the one induced by the slightly larger 
convective core due to the higher $\nabla_{r}$ (when a convective core 
is present).
Combining all those effects together, at the same evolutionary stage,
the solar-scaled  model 
is fainter, cooler and older than the \enh one.  
The same trend is recovered in the isochrones as well.

The implications of the different turn-off temperatures and 
luminosities on the ages of  globular clusters  have already been 
extensively investigated (see e.g.\ Salaris \& Weiss 1998).
For isochrones younger than $10^9$~yr, the differences
in the turn-off properties 
are much less marked. This is due both to (1) the weak metallicity
dependence
of the opacities  in massive stars 
(where electron-scattering becomes the main source of opacity) 
 and (2)  the larger extension of  central
convection caused by the flatter profile of $\nabla_{r}$ 
in  stellar interiors.

(ii) $\alpha$-Enhanced isochrones are hotter 
than the solar-scaled  ones throughout all evolutionary 
phases. The effect increases with the metallicity. It is  mainly due 
to the higher opacities of the solar-scaled  mixtures.
Looking at the  10~Gyr isochrone with chemical 
composition $[Z=0.040, Y=0.320]$ as an example, the temperature
difference $\log\Teff^{\rm solar} - \log\Teff^{\rm \alpha-enh}$ is about
$-$0.015 at the turn-off and increases to $-$0.020 at 
the bottom of the RGB. These differences in effective temperatures
result into
colour differences $(B-V)_{\rm solar} - (B-V)_{\rm \alpha-enh}$ of
0.05~mag and 0.08~mag, at the turn-off and bottom of the RGB, respectively.

\begin{table}
\caption{Position of the turn-off  in the HR diagram of isochrones with
metallicity  $Z=0.019$.}
\label{turnoff}
\begin{center}
\begin{tabular}{c|c|c|c|c}
\hline\noalign{\smallskip}
\multicolumn{1}{c}{}  & \multicolumn {2}{c}{solar-scaled } & 
	\multicolumn {2}{c}{$\alpha$-enhanced}  \\ 
        age    &   $\log(L/L_\odot)$    &  $\log T_{\rm eff}$  &      
  $\log(L/L_\odot)$    &  $\log T_{\rm eff}$  \\
\noalign{\smallskip}\hline\noalign{\smallskip}
 7.00  &    4.491 &  4.424 &  4.473 &  4.428 \\ 
 7.20  &    4.109 &  4.367 &  4.084 &  4.370 \\ 
 7.40  &    3.749 &  4.309 &  3.722 &  4.312 \\ 
 7.60  &    3.418 &  4.253 &  3.380 &  4.255 \\ 
 7.80  &    3.095 &  4.197 &  3.046 &  4.198 \\ 
 8.00  &    2.762 &  4.143 &  2.728 &  4.143 \\ 
 8.20  &    2.390 &  4.087 &  2.367 &  4.088 \\ 
 8.40  &    2.090 &  4.030 &  2.028 &  4.031 \\ 
 8.60  &    1.791 &  3.976 &  1.733 &  3.978 \\ 
 8.80  &    1.492 &  3.924 &  1.436 &  3.926 \\ 
 9.00  &    1.207 &  3.874 &  1.154 &  3.877 \\ 
 9.20  &    0.853 &  3.834 &  0.839 &  3.837 \\ 
 9.40  &    0.587 &  3.808 &  0.528 &  3.812 \\ 
 9.60  &    0.383 &  3.787 &  0.377 &  3.795 \\ 
 9.80  &    0.379 &  3.776 &  0.406 &  3.785 \\ 
 10.00 &    0.240 &  3.759 &  0.232 &  3.770 \\ 
 10.20 &    0.083 &  3.740 &  0.082 &  3.753 \\ 
\noalign{\smallskip}\hline\noalign{\smallskip}
\end{tabular}
\end{center}
\end{table}

It is worth remarking that in Weiss et al.\ (1995), the \enh
isochrones are found to have hotter and brighter turn-offs, 
and cooler RGBs (at least in large portions of it)  
than in the solar-scaled  case (see their Fig.~4; 
notice that the line-types there are 
wrong: solid lines refer to the \enh case). {The cooler RGBs 
are no longer found here or in VandenBerg et al.\ (2000) and
have been a consequence of the lack of low-temperature opacity 
tables for \enh compositions in opacities in Weiss et al.\ (1995).
This confirms how important is to 
use self-consistent opacities. }

\begin{figure}
\resizebox{\hsize}{!}{\includegraphics{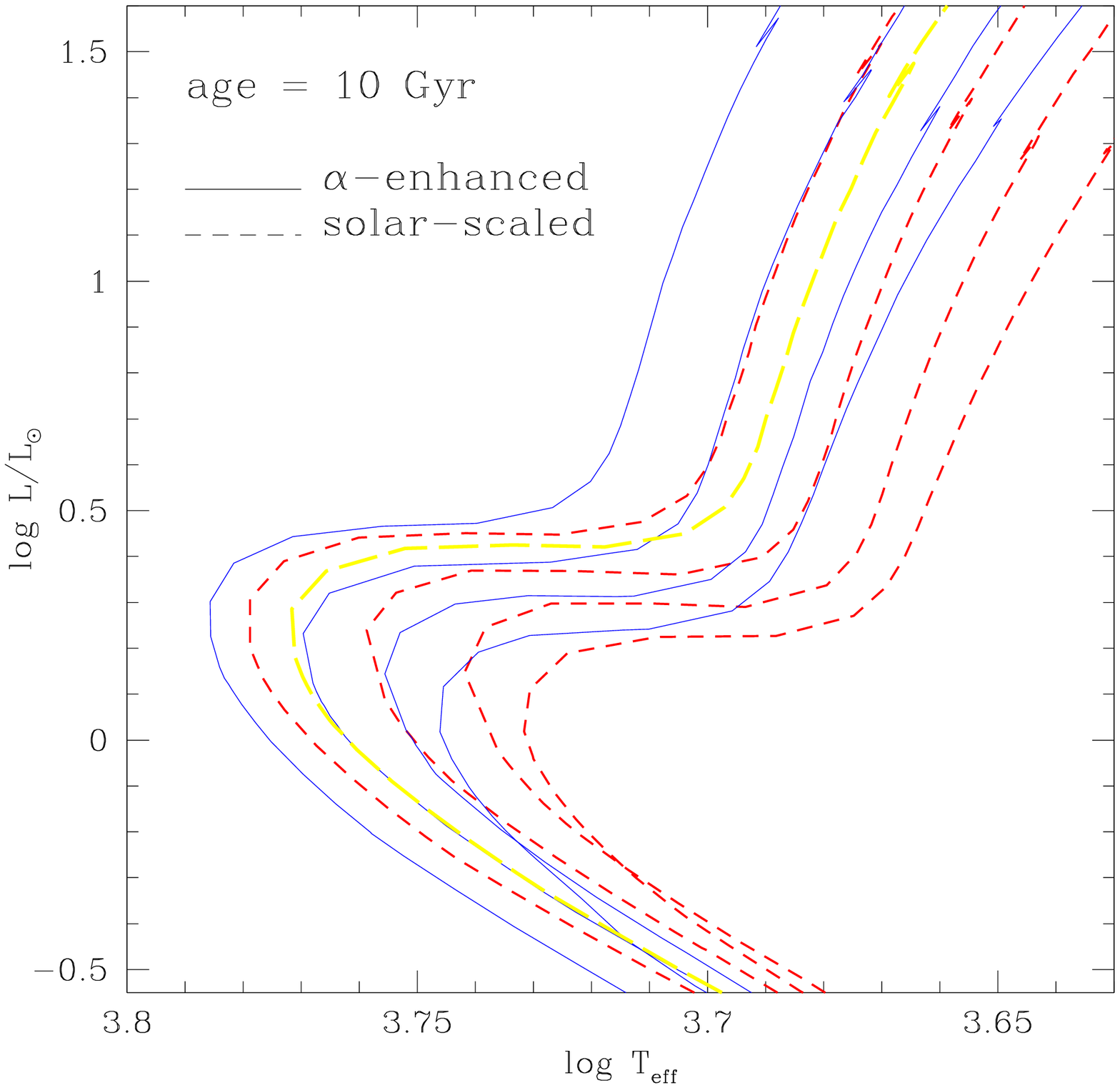}}
\caption{Comparison between \enh\ (continuous lines) 
and solar-scaled (short-dashed lines) isochrones of
10~Gyr. In both cases, from left to right the metallicity
values are $Z=0.008$, 0.019, 0.04, and 0.07.  
We also present a solar-scaled isochrone with $Z=0.011$ 
(long-dashed line), that has been obtained by interpolation 
among the solar-scaled tracks. This latter should be compared 
to the \enh\ isochrone with $Z=0.019$ (see text).}
\label{fig_isoc_10gyr}
\end{figure}
{
Finally, one can ask whether an \enh isochrone may be approximated by
a solar-scaled one with a slightly lower metallicity. In order
to answer this question let us examine a set of 10 Gyr old isochrones.
In Fig.~\ref{fig_isoc_10gyr} we plot all the isochrones with this age,
and with metallicity going from $Z=0.008$ to $Z=0.07$ for both the \enh 
(continuous lines) and solar-scaled (short dashed lines) mixtures.
One can see that both sets of isochrones 
have different shapes, the differences becoming larger at higher 
metallicities. More specifically, one can notice that
the difference in effective temperatures at the 
turn-off point between an \enh isochrone and 
a solar-scaled one, $\Delta\Teff(\rm TO)$, increases with the metallicity, 
passing from $\Delta\Teff(\rm TO)=0.008$ to $0.015$ as $Z$ increases 
from 0.008 to 0.07. 
On the contrary, the temperature difference at the base of the 
RGB, $\Delta\Teff(\rm RGB)$, is almost constant 
and of about $\Delta\Teff(\rm RGB)\sim 0.019$.

For the same age we 
calculate, by interpolating among the grids of different metallicity,
several solar-scaled isochrones with $Z$ spanning from $0.008$ 
to $0.019$, and selected among them the one best fitting the main 
sequence of the $Z=0.019$ \enh isochrone. This turns out 
to be the interpolated isochrone with $Z=0.011$, which is shown 
by the long-dashed line in Fig.~\ref{fig_isoc_10gyr}. Clearly, this 
solar-scaled $Z=0.011$ isochrone does not reproduce well the RGB 
sequence of the $Z=0.019$ \enh one. 
More specifically, although $\Delta\Teff(\rm TO)$ is 
very small for this isochrone pair, $\Delta\Teff(\rm RGB)$ 
is not neglegible ($\Delta\Teff(\rm RGB)\sim0.008$). 
 
From this simple test we then conclude that for a fixed age and 
relatively high metallicities, an \enh isochrone cannot be reproduced 
by tuning the metallicity of a solar-scaled one. 
}

The above temperature differences for RGB models (isochrones)  
bear very much on integrated spectral line indices of old metal-rich 
populations, which are commonly used to infer ages and metallicities 
of elliptical galaxies. This will be the subject of a forthcoming paper.

\subsection{Theoretical Johnson and HST photometry}

Theoretical luminosities and effective temperatures 
are  translated  to magnitudes and colours by means of  extensive
tabulations of bolometric
corrections (BC) and colours obtained from properly convolving 
spectral energy distributions (SEDs) as a function of $T_{\rm eff}$, 
$\log  g$ and [Fe/H].
The procedure is amply described in Bertelli et al. (1994), Bressan et al.
(1994), and 
Tantalo et al. (1996) to whom the reader should refer for more details.
Suffice it to recall here the following basic steps:

(i) The main body of the spectral library is from Kurucz (1992),
however extended to  the 
high and low temperature ranges.
For stars with $\Teff > 50,000$ K pure
black-body spectra are assigned, whereas for stars with $\Teff < 3500$
 K the catalogue of stellar fluxes by 
Fluks et al.\ (1994) is adopted.  This catalogue includes 97
observed spectra for all M-spectral subtypes in the wavelength range 
$3800 \le \mbox{\AA} \le 9000$, and synthetic photospheric spectra 
in the range $9900 \le \mbox{\AA} \le 12500$. 

The scale of \Teff\ in Fluks et al.\ (1994) is similar
to that of Ridgway et al.\ (1980) for spectral types earlier than M4 but
deviates from it for later spectral types. 
Since Ridgway's et al.\ (1980)
scale does not go beyond the spectral type M6, no comparison for more
advanced spectral types is  possible. 

The problem is further complicated by possible effects of metallicity. The
Ridgway scale of \Teff\ is based on stars with solar metallicity 
($Z \sim 0.02$) and empirical  calibrations of the  
\Teff-scale for $Z \neq 0.02$ are not available.

To cope with this difficulty, we have introduced the 
metallicity-\Teff\ relation of Bessell et al.\
(1989, 1991) using the ($V-K$) colour as a temperature
indicator. An interpolation is made between the \Teff\ of 
Bessell et al.\ (1989)
and the ($V-K$) colours given by Fluks et al.\ (1994) for the spectral
types from M0 to M10. 

(ii) {At a given metallicity} \feh, surface gravity $\log g$, and 
\Teff, BCs are determined by convolving the corresponding stellar 
SED  with the response 
functions of the various pass-bands. 

(iii) We recall that Kurucz does not provide atmospheric models for both 
solar-scaled and \enh mixtures.
It is not therefore possible to estimate the impact (if any) 
of \enh atmospheres  on the isochrone colours and magnitudes.

\paragraph{Johnson-Cousins photometry.}

The  response functions for the various pass-bands in which magnitudes and 
colours are generated are from the following sources:
 Buser \& Kurucz (1978) for the $UBV$ pass-bands, 
 Bessell (1990)  for the  $R$ and $I$ Cousins pass-bands, and finally 
  Bessell \& Brett (1988) for the $JHK$ pass-bands.
  
Re-normalization  of the colours obtained by convolving the SEDs with
pass-band, has been made by convolving the SED of  Vega 
and imposing
that the computed colours  strictly  match the observed ones (Kurucz 1992).

Finally, the zero  point of the BCs is fixed by imposing that the BC 
for the Kurucz model of the Sun is $-0.08$.  

\paragraph{HST-WFPC2 photometry.}

The transformation to the WFPC2 photometric system 
(filters F170W, F218W, F255W, F300W, F336W, F439W, F450W, F555W, 
F606W, F702W, F814W, and F850LP) requires some additional explanations.

The main parameters defining this photometric system are given 
in Table~\ref{hstfilter}.  
Columns (1) to (3) report the filter name, 
the mean wavelength $\bar{\lambda}$ and the 
r.m.s.\ band width (BW), respectively. 

\begin{table*}
\caption{Main characteristics of the WFPC2-HST photometric 
system and zero points.}
\label{hstfilter}
\begin{center}
\begin{tabular}{c|c|c|c|c|c}
\noalign{\smallskip}\hline\noalign{\smallskip}
 Filter  & $\bar{\lambda}$ (\AA) & BW (\AA) & $Z_P$ & 
$m_{\rm ST} - m_{\rm VEGA}$ & Note \\
(1) & (2) & (3) & (4) & (5) & (6)  \\
\noalign{\smallskip}\hline\noalign{\smallskip}
\hline
          F170W   & 1747   &  290   & $-$20.6926    & $-$0.4070 &  $-$ \\
          F218W   & 2189   &  171   & $-$20.8737    & $-$0.2263 &  $-$ \\
          F255W   & 2587   &  170   & $-$21.0940    & $-$0.0060 &  $-$ \\
          F300W   & 2942   &  325   & $-$21.1336    & +0.0336 & Wide U \\
          F336W   & 3341   &  204   & $-$21.1876    & +0.0876 & U-John. \\
          F439W   & 4300   &  202   & $-$20.4325    & $-$0.6675 & B-John. \\
          F450W   & 4519   &  404   & $-$20.6339    & $-$0.4661 & Wide B \\
          F555W   & 5397   &  522   & $-$21.0798    & $-$0.0202 & V-John. \\
          F606W   & 5934   &  637   & $-$21.4250    & +0.3250 & Wide V \\
          F702W   & 6862   &  587   & $-$21.8594    & +0.7594 & Wide R \\
          F814W   & 7924   &  647   & $-$22.3309    & +1.2309 & I-John. \\
          F850LP  & 9070   &  434   & $-$22.7578    & +1.6578 & $-$ \\
\noalign{\smallskip}\hline\noalign{\smallskip}
\end{tabular}
\end{center}
\end{table*}

With the WFPC2 detector  at least three kinds of magnitudes are 
commonly in use: 
$m_{\rm ST}$, $m_{\rm VEGA}$ and $m_{\rm AB}$ 
(see the {\it Synphot User's Guide} by  White et al. 1998, distributed 
by the STSDAS Group). 
In the following we limit ourselves to consider  only 
the theoretical counterparts 
of $m_{\rm ST}$ and $m_{\rm VEGA}$.

For radiation with  flat spectrum and specific flux $F_P$ 
(in ${\rm erg\,\,s^{-1}\,cm^{-2}\,\mbox{\AA}^{-1}}$) impinging on   the
telescope, the $m_{\rm ST}$ magnitude is defined by
\begin{equation}
m_{{\rm ST}, P} = -21.1 -2.5 \log F_P 
\label{stmagdef}
\end{equation}
which means that $m_{{\rm ST},P}=0$ for 
$F_P = 3.63\,10^{-9}$ $\rm erg\,\,s^{-1}\,cm^{-2}\,\mbox{\AA}^{-1}$.

Let us then define the pass-band $P(\lambda)$
as  the product of the filter transmission by the response function 
of the telescope assembly and detector in use. In this case,
$F_P$ is related to the {\it counts} rate $\dot C_P$ -- i.e.\ the number
of photons per second registered by the detector -- through
\begin{equation}
F_P = \frac{hc}{A \int P(\lambda) \lambda \,\diff\lambda} \,\dot C_P
\label{countrate}
\end{equation}
where $A$ is the effective collecting surface of the 
telescope, and $h$ and $c$ are the Planck constant and speed of light, 
respectively.

$\dot C_P$,  the quantity actually measured at the telescope, can be  easily 
converted to a specific flux by means of Eq.~(\ref{countrate}). Therefore, 
specific fluxes $F_P$  and $m_{{\rm ST}, P}$ magnitudes can be defined 
for incoming spectra of any shape. In the case of a stellar source of radius
$R$, located at a distance $d$, and emitting a specific flux $I_{\rm K}$,
Eq.~(\ref{stmagdef}) can be replaced by 
\begin{equation}
m_{{\rm ST}, P} = -21.1 -2.5 \log\left[ \frac{R^2}{d^2} 
\frac{\int P(\lambda) \,I_{\rm K}(\lambda) \,\lambda \,\diff\lambda}
{\int P(\lambda) \,\lambda \,\diff\lambda} \right]  
\label{stmagdefstar}
\end{equation}
which  defines the apparent $m_{{\rm ST}, P}$ 
magnitude of a star. 

In our case, we derive the absolute $M_{{\rm ST}, P}$ magnitudes by simply 
locating our synthetic stars at a distance of $d=10$~pc in 
Eq.~(\ref{stmagdefstar}). 
The specific fluxes $I_{\rm K}(\lambda)$ are taken from the library of
synthetic spectra in use as a function of 
$T_{\rm eff}$, surface gravity $g$, and chemical composition.

Similarly, we can define the $m_{{\rm VEGA}, P}$ magnitudes
\begin{equation}
m_{{\rm VEGA}, P} = Z_P -2.5 \log\left[ \frac{R^2}{d^2} 
\frac{\int P(\lambda) \,I_{\rm K}(\lambda) \,\lambda \,\diff\lambda}
{\int P(\lambda) \,\lambda \,\diff\lambda} \right]  
\label{vegamagdefstar}
\end{equation}
by imposing that the zero points $Z_P$ are such that 
the synthetic magnitudes of Vega match the ground-based $UBVRI$ 
Johnson apparent magnitudes for this star,
for the filters which are closest in wavelength to each other 
(see column 6 in Table~\ref{hstfilter}). 

We adopt the values
0.02, 0.02, 0.03, 0.039, 0.035~mag for the apparent magnitudes of Vega 
in the $UBVRI$ Johnson system (see Holtzman et al.\ 1995).
For the UV filters, Vega is assumed to have an apparent magnitude equal 
to zero. 

We remind the reader  that if we  assume the apparent Vega magnitude 
to be equal to zero in
{\em all} the pass-bands, we would obtain 
$Z_P=2.5 \log F_{P, {\rm Vega}}$ and Eq.~(\ref{vegamagdefstar}) would become 
\begin{equation}
m_{{\rm VEGA},P} = -2.5 \log \frac{F_P}{F_{P,{\rm Vega}}} 
\end{equation}
This latter is the definition of {\em vegamag} in the SYNPHOT package 
distributed with the STSDAS software. 
In this paper we do not make use of this calibration, but prefer 
to follow the slightly
different one defined by Holtzman et al.\ (1995) and described above. 

To get the final  calibration, we insert in Eq.~\ref{vegamagdefstar} 
the  specific flux of Vega calculated by Castelli \& Kurucz (1994) with 
atmosphere models, 
and converted it to an absolute flux on the Earth assuming for 
Vega the angular diameter of $3.24$~mas (Code et al.\ 1976). 
The absolute flux we have adopted differs (a few  
percent) from the one by Hayes (1985)  used by Holtzman et al.\ (1995). 
Once more, the absolute magnitude $M_{{\rm VEGA},P}$ is obtained locating the 
synthetic star at a 10~pc distance. Column (4) of Table
\ref{hstfilter} lists the zero points  we have obtained, whereas column (5) 
gives the difference $-21.1 - Z_P$ between  the zero point of the
$m_{\rm ST}$  and $m_{\rm VEGA}$ magnitudes. The conversion from one magnitude
scale to another is thus possible.

Finally, we  calculate the bolometric corrections BC which allow one 
to convert theoretical  bolometric magnitudes 
both to $M_{\rm ST}$ and $M_{\rm VEGA}$:
\begin{eqnarray}
{\rm BC}(P)^{M_{\rm ST}}_{T_{\rm eff},g} & = & M_{\rm bol}^{\odot} -2.5 
\log\left[ \frac{4 \pi (10 {\rm pc})^2 \sigma T_{\rm eff}^4}{L_{\odot}} \right]
\nonumber \\
  & + & 21.1  + 2.5 \log \frac{\int P(\lambda) \,I_{\rm K}(\lambda)\, \lambda 
\,\diff\lambda}{\int P(\lambda)\, \lambda \,\diff\lambda} \\
{\rm BC}(P)^{M_{\rm VEGA}}_{T_{\rm eff},g} & = &
{\rm BC}(P)^{M_{\rm ST}}_{T_{\rm eff},g} -21.1 - Z_P
\label{bc_stmag}
\end{eqnarray}
where $\sigma$ is the Stefan-Boltzmann constant.  For the  solar luminosity
and absolute bolometric magnitude we adopt
 $L_\odot=3.844~10^{33}~\rm erg~s^{-1}$ (Bahcall et al.\ 1995)
 and $M_{\rm bol}^{\odot}=4.77$.

Owing to the presence of contaminants inside the  WFPC2 
(see Baggett \& Gonzaga  1998
and Holtzman et al.\ 1995), $P(\lambda)$
changes slowly with time; in particular the UV throughput degrades, 
changing the photometric performances. To cope with this 
drawback, small corrections are usually added
 to the definition of the instrumental magnitudes  
in order to bring the magnitudes back to the optimal conditions
(see Holtzman et al.\ 1995). Because of this,  in the calculation of 
the bolometric corrections 
we do not use the present-day pass-bands $P(\lambda)$ provided by
SYNPHOT but insert  the pre-launch pass-bands, as in 
Holtzman et al.\ (1995).

\subsection{Integrated magnitudes and mass-to-light ratios of single stellar
populations}

In this section we present the  integrated magnitudes, colours  and 
mass-to-light ratios  for single stellar populations (SSP) of the 
same age and chemical compositions of the isochrones above, 
both in the Johnson-Cousins and WFPC2
photometric systems. 

The integrated magnitudes and colours are computed assuming  an
initial mass function  $\phi(M)$ and adopting the normalization
\begin{equation}
\int_{M_{\rm l}}^{M_{\rm u}} M \,\phi(M)\,\diff M = 1~M_{\odot}
\label{eq_imf}
\end{equation}
where $M_{\rm l}$ and $M_{\rm u}$ are the lower and upper limits of 
zero age main sequence stars. By doing so, the integrated magnitudes 
of these ideal SSPs can be easily scaled to 
produce integrated magnitudes for stellar populations
of any arbitrary total {\em initial} mass $M_{\rm T,i}$, by simply
adding the factor $-2.5\log M_{\rm T,i}$. 

The theoretical mass-to-light
ratios $M/L$ are calculated  following the same procedure as in 
Alongi \& Chiosi (1990) and Chandar et al.\ (1999).

The fate of a single star depends on its initial mass (binary stars 
are neglected here). In a brief and over-simplified
picture of stellar evolution the following mass limits and groupings
can be identified: (i) Stars more massive
than {$M_{\rm H}\simeq 0.7-0.8~M_{\odot}$ }
have a lifetime shorter than the current estimate of the Hubble age
of the Universe, say $13-15$ Gyr. Stars lighter than this limit are not of
interest here because once formed they live for ever. 
(ii) Stars more massive than $M_{\rm up}$
explode as supernovae leaving a neutron star remnant of $1.4\, M_{\odot}$.
The possibility that massive stars ($M \ge 25 \, M_{\odot}$) may end up
as Black Holes is not considered here. (iii) Stars less massive
than $M_{\rm W}$ and more massive than 
$M_{\rm H}$ terminate their evolution as White Dwarfs of suitable 
{masses} $M_{\rm WD}$ that depend on the initial 
mass and efficiency of mass loss during the RGB
and AGB phases. With the current estimates of the mass loss
efficiency $M_{\rm W}\simeq 5-6 \, M_{\odot}$, i.e.\ 
$M_{\rm W}\simeq M_{\rm up}$.
The possibility, however, exists that $M_{\rm W} < M_{\rm up}$. In such a case
stars in the mass interval $M_{\rm W}$ to $M_{\rm up}$ reach the C-ignition
stage and deflagrate as supernovae leaving no remnant. For the purposes
of the present study, this latter case is neglected and $M_{\rm W}=M_{\rm up}$
is always assumed. Finally, let $M_{\rm a}$ be the  
initial mass of {the most} massive star still alive in a SSP 
of a certain age. It is worth recalling that all the above mass
limits depend on the initial chemical composition and that this
dependence can be properly taken into account.

{
Therefore, at any given age $t$ the total mass $M_{\rm T}$ of a SSP 
is made of two contributions:
\begin{equation}
       M_{\rm T}(t) = \int_{M_{\rm l}}^{M_{\rm a}}
	M\,\phi(M)\,\diff M +
	\int_{M_{\rm a}}^{M_{\rm u}}
	M_{\rm r}\,\phi(M)\,\diff M
\end{equation}
where the first right-hand side term 
is the total mass of the stars still alive 
at the age $t$ (i.e.\ burning a nuclear fuel) and the
second one is the total mass  in  stellar remnants, i.e.\  
white dwarfs or neutron stars depending on the initial mass 
of the progenitor (i.e.\ $M_{\rm r}(M)=M_{\rm WD}(M)$ for
$M\le M_{\rm up}$, and $M_{\rm r}(M)=1.4~M_\odot$ for
$M > M_{\rm up}$). 
We adopt an upper mass limit $M_{\rm u}=120~M_{\odot}$. 
The choice for the lower mass limit, $M_{\rm l}$, 
is explained below. It is worth recalling
here that, due to eq.~\ref{eq_imf}, 
$M_{\rm T}(t)$ is the current mass of an ideal SSP with 
initial mass equal to $1~M_\odot$. $M_{\rm T}(t)$
decreases as a function of time from its initial value, 
due to the mass lost by both stellar winds (for all masses) 
and supernova explosions (from massive stars). 
}

Finally, the total luminosity $L_j$ of the SSP in any pass-band $j$ 
is obtained by integrating, along the 
isochrone, the luminosity of individual stars whose number
is {given by} the IMF 
\begin{equation}
L_j = \int_{M_{\rm l}}^{M_{\rm a}} L_j(M)\, \phi(M)\, \diff M ~.
\end{equation}

To proceed further, one has to specify the initial mass function
and $M_{\rm l}$.
Here, we adopt the Salpeter (1955) law with slope 
$x=1.35$ over the whole mass range  $M_{\rm l}$  to $M_{\rm u}$.
Concerning $M_{\rm l}$, the choice $M_{\rm l}=0.068~M_{\odot}$ 
is made by imposing 
that the observational value  $M/L_V=0.19 \pm 0.04$ of the LMC cluster 
NGC~1866 is matched (see Girardi \& Bica 1993).
The age and metallicity of NGC~1866 are $10^{8}$ years and $Z=0.008$,
respectively, whereas its $M/L_V$ is close to the mean value
of 0.20 for LMC clusters of the same age
(see Battinelli \& Capuzzo-Dolcetta 1989, and 
references therein).

\section {Concluding remarks}
\label{sec_conclusion}

We have computed extended sets of \enh  evolutionary tracks 
and isochrones at relatively high metallicities. A major improvement,
with respect to previous calculations of \enh tracks, is that 
we made  use of self-consistent  opacities, i.e.\ computed with the
same \enh chemical compositions over the complete  range of
temperatures. The main result is that in general  all
evolutionary phases (isochrones) have higher effective temperature with
respect to solar-scaled  models (isochrones) of same metal content $Z$.
The temperature shift is  caused by the lower
opacities  of the  \enh mixtures. {Moreover, we find that 
at relatively high metallicities and old ages, an \enh isochrone 
cannot be mimicked by simply using a solar-scaled isochrone of
lower metallicity.}

Obvious limitation of the present results is that they 
refer to a particular choice for the \enh chemical
compositions derived from   observations of a particular sample
of metal-poor field stars. 
Future observational data may suggest a different partition of metals
in  \enh stars, and hence different results for the
stellar models and corresponding  isochrones. 

Testing  a large
number of possible combinations of \enh ratios, though feasible,
would not be of practical use owing to the present uncertainty on the
abundance ratios. Therefore, the results of this study ought to be
taken as indicative of the {\em potential}
effect of \enh element ratios.

Moreover, it is likely that  variations of single $\alpha$-elements 
are less important than  the whole problem of including or not 
the enhancement of $\alpha$-elements in the initial chemical composition. 
Finally, it is still not clear whether
differences in enhancements are significant compared to the errors
in the determinations.

\paragraph{Retrieval of the data sets.}
Complete tabulations of the relevant data for all 
evolutionary sequences, isochrones, integrated magnitudes and colours
and mass-to-light rations  together with useful summary tables can be 
obtained either upon request to the authors or 
downloaded from the address \verb$http://pleiadi.pd.astro.it$.

The layout of the tables of stellar models and isochrones
is the same as in Girardi et al.\ (2000) 
as far as the 
Johnson-Cousins photometric system is concerned. For the WFPC2 
system the layout is similar, with table headers allowing to easily 
identify the pass-band.

\vspace{0.5truecm}
{\it Acknowledgments.}
We like to thank A.\ Bressan for many  suggestions and useful
conversations,  H.\ Schlattl for his help with the opacity interpolation code, 
H.\ Aussel and  M.\ Zoccali for their useful clarifications about the 
HST/WFPC2 photometric system, J.\ Holtzman for providing us with 
the pre-launch WFPC2 pass-bands, and the anonymous referee for useful
suggestions.
L.G.\ acknowledges support from the Alexander von Humboldt-Stiftung 
during his stay at MPA. B.S.\ thanks MPA for the warm hospitality and
support.
This study has been funded by the Italian
Ministry of University, Scientific Research and Technology  
(MURST) under contract  ``Formation and Evolution of Galaxies'' n.\ 
9802192401.


\section*{References}

\begin{description}

\item Alexander D.R., Ferguson J.W., 1994, ApJ 437, 879
\item Alongi M., Chiosi C., 1990, in Astrophysical Ages and Dating Methods, 
      ed.\ E.\ Vangioni-Flam et al.\ 
      (Gif-sur-Yvette: Editions Fronti\`eres), 207
\item Alongi M., Bertelli G., Bressan A., Chiosi C., 1991, A\&A
    	244, 95
\item Anders E., Grevesse N., 1989, Geochim. Cosmochim. Acta, 53,197
\item Baggett S., Gonzaga S., 1998, ISR WFPC2 98-03
\item Bahcall J.N., Pinsonneault M.H., Wasserburg G.J., 1995, 
	Rev.\ Mod.\ Phys.\ 67, n.4, 781 
\item Battinelli P., Capuzzo-Dolcetta R., 1989, ApJ 347, 794
\item Bertelli G., Bressan A., Chiosi C., Fagotto F., Nasi E., 1994, 
	A\&AS 106, 275
\item Bessell M.S., 1990, PASP 102, 1181
\item Bessell M.S., Brett J.M., 1988, PASP 100, 1134
\item Bessell M.S., Brett M.J., Wood P.R., Scholz M., 1989,  A\&AS 77, 1
\item Bessell M.S., Brett M.J., Wood P.R., Scholz M., 1991,  A\&AS 87, 621
\item B\"ohm-Vitense E., 1958, Z. Astroph. 46, 108  
\item Bressan A., Bertelli G., Chiosi C., 1981, A\&A 102, 25
\item Bressan A., Fagotto F., Bertelli G., Chiosi C., 1993,
        A\&AS 100, 647
\item Bressan A., Chiosi C., Fagotto F., 1994, ApJS 94, 63
\item Buser R., Kurucz R.L. 1978, A\&A 70, 555
\item Carney B.W., 1996, PASP, 108, 900
\item Catelan M., De Freitas Pacheco J.A., Horvath J.E., 1996, ApJ, 461, 231
\item Cassisi S., Castellani V., Degl' Innocenti S., Weiss A., 1998, 
	 A\&AS 129, 267
\item Castelli F., Kurucz R.L., 1994, A\&A 281, 817
\item Castellani V., Degl'Innocenti S., Girardi L., et al., 2000, A\&A 354, 150
\item Caughlan G.R., Fowler W.A., 1988, Atomic Data Nucl.\ Data Tables
        40, 283
\item Chandar R., Bianchi L., Ford H.C., Salasnich B., 1999, PASP 111, 794
\item Chiosi C., Vallenari A., Bressan A., 1997, A\&AS 121, 301
\item Code A.D., Bless R.C., Davis J., Brown R.H., 1976, ApJ, 203, 417
\item Fagotto F., Bressan A., Bertelli G.P., Chiosi C., 1994, A\&AS 105, 29
\item Fluks M.A, Plez B., The P.S., et al., 1994, A\&AS 105, 311
\item Girardi L., 1999, MNRAS 308, 818  
\item Girardi L., Bica E., 1993, A\&A 274, 279  
\item Girardi L., Bertelli G., 1998, MNRAS 300, 533
\item Girardi L., Bressan A., Bertelli G., Chiosi C., 2000, A\&AS  141, 371
\item Grevesse N., Noels A., 1993, Phys.\ Scr.\ T, 47, 133   
\item Haft M., Raffelt G., Weiss A., 1994, ApJ 425, 222
\item Hayes D.S., 1985, Calibration of fundamental stellar quantities, 
  	IAU Symposium 111, ed.\ D.S.\ Hayes, L.E.\ Pasinetti and A.G.D.\ 
	Philip (Dordrecht, Reidel), p.\ 225
\item Holtzman J.A., Burrows C.J., Casertano S., et al., 1995, PASP, 107, 1065
\item Hubbard W.B., Lampe M., 1969, ApJS 18, 297
\item Iglesias C. A., Rogers F. J., 1993, ApJ 412, 752
\item Iglesias C. A., Rogers F. J., 1996, ApJ 464, 943
\item Itoh N., Mitake S., Iyetomi H., Ichimaru S., 1983, ApJ 273, 774
\item King J., 1994, PASP, 106,423
\item Kurucz R.L., 1992, in  IAU
     	Symp. 149: The Stellar Populations of Galaxies, 
	eds.\ B.\ Barbuy, A.\ Renzini, Dordrecht, Kluwer, p.\ 225   
\item Landr\'e V., Prantzos N., Aguer P., et al., 1990, A\&A 240, 85
\item Matteucci F., Brocato E., 1990, ApJ 365, 539
\item Matteucci F., Romano D., Molaro P., 1999, A\&A 341, 458
\item McWilliam A., Rich R.M., 1994, ApJS, 91,749
\item Mihalas D., Hummer D.G., Mihalas B.W., D\"appen W., 1990, ApJ 350, 300
\item Munakata H., Kohyama Y., Itoh N., 1985, ApJ 296, 197
\item Rich R.M., McWilliam A., 2000, in Discoveries and Research Prospects
     from 8--10-Meter-Class Telescopes, Proc.\ of the SPIE vol.\ 4005, 
	ed.\ J.\ Bergeron, in press (astro-ph/0005113)  
\item Ridgway S.T., Joyce R.R., White N.M., Wing R.F., 1980, ApJ 235, 126
\item Rogers F.J., Iglesias C.A., 1992, ApJS 79, 507
\item Rogers F.J., Iglesias C.A., 1995, in ASP Conf.\ Proc.\ 78, 
      	Astrophysical Applications of Powerful New Databases, ed.\ 
      	S.J.\ Adelman, W.L.\ Wiesse (San Francisco:ASP), 31
\item Ryan S.G., Norris J.E., Bessel M.S., 1991, AJ 102, 303
\item Salaris M., Weiss A., 1998, A\&A 335, 943
\item Salaris M., Chieffi A., Straniero O., 1993, ApJ 414, 580
\item Salaris M., Degl'Innocenti S., Weiss A., 1997, ApJ 479, 665
\item Salpeter E.E, 1955, ApJ 121, 161
\item Schlattl H., Weiss A.,  1998, in "Proc. Neutrino Astrophysics" , 
	Ringberg Castle, Tegernsee, Germany, 
        20-24 Oct 1997, ed.\ M.\ Altmann, W.\ Hillebrandt, 
	H.-T.Janka and G.Raffelt 
        (SFB Astroteilchenphysik, Technical University Munich)
\item Sp\"ath H., 1973, in "Spline-Algorithmen zur Konstruktion 
	glatter Kurven und Fl\"achen", M\"unchen: Oldenbourg.
\item Tantalo R., Chiosi C., Bressan A., Fagotto F., 1996, A\&A 311, 361
\item VandenBerg D.A., Swenson F.J., Rogers F.J., Iglesias C.A., 
	Alexander D.R., 2000, ApJ 532, 430 
\item Weiss A., Keady J.J., Magee N.H.~Jr., 1990, 
	Atomic Data and Nuclear Data Tables 45, 209
\item Weiss A., Peletier R.F., Matteucci F., 1995, A\&AS 296, 73
\item White R., Greenfield P., Kinney E. et al., 1998, ed.\ by Association of 
      Universities for Research in Astronomy, Inc.
\item Worthey G., Faber S.M., Jes\'us Gonz\'alez J., 1992, ApJ 398, 69

\end{description}  

\end{document}